\documentclass[Conference,a4paper]{IEEEtran}
\IEEEoverridecommandlockouts
\usepackage{color}
\usepackage{graphicx}
\usepackage{epstopdf}
\usepackage{amsmath}
\usepackage{amssymb}
\usepackage{algorithm}
\usepackage{algorithmic}
\usepackage{amsmath}
\usepackage{multirow}
\usepackage{booktabs}
\usepackage{array}
\usepackage{amsthm}
\usepackage{stfloats}
\usepackage{caption}
\usepackage{subfigure}
\usepackage{bm}
\usepackage{booktabs}
\usepackage{setspace}
\usepackage{diagbox}
\usepackage{enumerate}
\usepackage{ulem}
\usepackage{url}
\usepackage{circledsteps}

\usepackage{hyperref}

\allowdisplaybreaks[4]

{\bgroup
 \addtolength\abovedisplayshortskip{#1}
 \addtolength\abovedisplayskip{#1}
 \addtolength\belowdisplayshortskip{#1}
 \addtolength\belowdisplayskip{#1}
 }
{\egroup\ignorespacesafterend}

\newtheorem{theorem}{Theorem}
\newtheorem{lemma}{Lemma}

\newcommand{\be}{\begin{equation}}
\newcommand{\ee}{\end{equation}}
\newcommand{\bea}{\begin{eqnarray}}
\newcommand{\eea}{\end{eqnarray}}
\newcommand{\ba}{\begin{array}}
\newcommand{\ea}{\end{array}}

\title{{ Cram\'er-Rao Bound Optimization for Fluid Antenna-Empowered Integrated Sensing and Uplink Communication System}}
\author{\IEEEauthorblockN{Yuan Guo, Wen Chen, Qingqing Wu, Yang Liu, and Qiong Wu
\thanks{
Y. Guo, W. Chen and
Q. Wu are with Department of Electronic Engineering, Shanghai Jiao Tong University, Shanghai, China, 
email:
yuanguo26@sjtu.edu.cn,
wenchen@sjtu.edu.cn,
qingqingwu@sjtu.edu.cn.}
\thanks{
Y. Liu  is with the School of Information
and Communication Engineering, Dalian University of Technology, Dalian, China, 
email:
yangliu\_613@dlut.edu.cn.}
\thanks{
Q. Wu is with the School of Internet of Things Engineering, 
Jiangnan University, Wuxi, China, 
emali: qiongwu@jiangnan.edu.cn.
}
}
}

\begin{document}
\maketitle
\pagestyle{empty}
\thispagestyle{empty}

\begin{abstract}

Integrated sensing and communication (ISAC) is a promising solution for the future sixth-generation (6G) system.
However, classical fixed-position antenna (FPA) ISAC systems fail to fully utilize spatial degrees of freedom (DoFs), 
resulting in limited gains for both radar sensing and communication functionalities. {
This challenge can be addressed by the emerging novel fluid antenna (FA) technology,}
which can pursue better channel conditions and improve sensing and communication performances.
In this paper, we aim to minimize the Cram$\text{\'{e}}$r-Rao bound (CRB) for estimating the target's angle while guaranteeing communication performance. This involves jointly optimizing active beamforming, power allocation, receiving filters, and FA position configurations, which is a highly non-convex problem.
To tackle this difficulty, 
we propose an efficient iterative solution that analytically optimizes all variables without relying on numerical solvers, i.e., CVX. 
Specifically, by leveraging cutting-edge majorization-minimization (MM) and penalty-dual-decomposition (PDD) methods, we develop a low-complexity algorithm to solve the  beamformer configuration problem containing the fractional and quartic terms.
Numerical simulation results demonstrate the effectiveness and efficiency of our proposed algorithm, highlighting significant performance improvements achieved by employing FA in the ISAC system.

\end{abstract}


\maketitle
\section{Introduction}

Recently, 
with the development of emerging sixth-generation (6G) applications, including smart homes, 
vehicle-to-everything (V2X) communications, 
and environmental monitoring, 
the demand for high-quality communication and precise sensing capabilities has been steadily increasing. In this context, integrated sensing and communication (ISAC) is widely regarded as a promising technique to meet these demands through the joint design of communication and sensing functions \cite{ref_ISAC_LF}. 
It has attracted significant interest from both academia and industry. 
By sharing spectrum resources and utilizing a unified hardware platform, 
ISAC aims to significantly enhance spectrum efficiency and reduce hardware costs and system complexity. 
Furthermore, many recent studies that extensively document dual-functional waveform designs for joint radar sensing and communication can be found in \cite{ref_ISAC_LF}$-$\cite{ref_ISAC_meng}, along with their references.

Nonetheless, 
traditional ISAC systems equipped with fixed-position antennas (FPAs) cannot fully exploit spatial diversity, 
which results in a loss of beamforming gains for both communication and sensing tasks. 
To address this limitation,
an innovative concept is fluid antenna system (FAS) was proposed in \cite{FAS_1}$-$\cite{FAS_2}.
This novel FAS architecture which utilizes flexible antenna technology such as
liquid-based antenna \cite{FAS_3}$-$\cite{FAS_4},
mechanically movable antenna \cite{ref_channel model_1}$-$\cite{ref_channel model_2},
and etc.,
to adjust antenna's position flexibly within a predefined spatial area.
By fully leveraging the additional degrees of freedom (DoFs) provided by the reconfiguration of wireless channels, 
the FA structure aims to enhance communication capabilities \cite{ref_channel model_1}$-$\cite{ref_channel model_2}. 
The significant potential of the FA technology to improve communication performance has been extensively demonstrated in recent studies 
\cite{ref_channel model_1}$-$\cite{ref_MA_mag}, 
including 
uplink communication \cite{ref_channel model_2}, \cite{ref_MA_communication_0},
interference network \cite{ref_MA_communication_1},
nonorthogonal multiple access (NOMA) \cite{ref_MA_communication_2},
multicast communication \cite{ref_MA_communication_3},
along with their references.

\subsection{Related Works}
Due to the significant benefits of FA technique,
extensive research focused on its integration into ISAC systems to substantially enhance both communication capacity and sensing ability, e.g., \cite{ref_related_work_new_1}$-$\cite{ref_related_work_new_5}.
The literature \cite{ref_related_work_new_1} designed a port selection strategy for the FA system to reduce transmit power.
A  FA-aided ISAC system was researched in \cite{ref_related_work_new_2}.
The study aimed to maximize downlink (DL) communication rate 
while satisfying the sensing beampattern gain requirements.
The authors of \cite{ref_related_work_1}
first considered deploying the FA into 
unmanned aerial vehicle (UAV)-enabled ISAC system
to support low-altitude economy (LAE) applications.
Furthermore, 
an efficient algorithm,
which jointly optimizes beamforming and the positions of the FA,
has been proposed to promote both throughput capacity and beamforming gain.
The work \cite{ref_related_work_2} focused on minimizing the Cram\'er-Rao bound (CRB) for estimating the direction of arrival (DoA) of a target in a DL communication ISAC system assisted by FA technology,
and demonstrated that FA technology can significantly improve sensing performance.
The paper \cite{ref_related_work_3} employed FA technology
to enhance both communication and sensing performances 
in a reconfigurable intelligent surface (RIS)-assisted ISAC system.
Besides,
the numerical results demonstrated that FA efficiently reduces multiuser interference 
and mitigates the impact of multi-path effects.
The authors of \cite{ref_related_work_new_3} adopted the deep reinforcement learning framework
to jointly optimize the antenna port locations and precoding design in a multiuser 
multiple-input multiple-output (MIMO) downlink ISAC system.
In \cite{ref_related_work_4},
under the impact of the nature of the dynamic radar cross-section (RCS),
the authors investigated the transmit power minimization problem 
while assuring the individual quality-of-service (QoS) requirement
for communication and sensing in an FA-aided ISAC system.
The authors of \cite{ref_related_work_6} adopted the emerging FA architecture in the ISAC system with low-altitude airborne vehicles
and showed it could remarkably improve the communication capacity under the constraint of sensing signal-to-noise ratio (SNR).
The work \cite{ref_related_work_7} firstly deployed the FA into a bi-static radar system
with the objective of maximizing both the weighted communication rate and sensing mutual information (MI). 
Furthermore, 
the positions of the FA can be analytically
updated by leveraging the Karush-Kuhn-Tucker (KKT) conditions.
The authors of \cite{ref_related_work_7_1} aim to maximize the sensing signal-to-interference-plus-noise ratio (SINR) 
while assuring the minimum SINR of mobile users in an FA-assisted bi-static ISAC system.
The paper \cite{ref_related_work_8} employed the FA to guard the communication security of RIS-aided ISAC network effectively, 
and proposed a two-layer penalty-based algorithm to solve the non-convex communication rate maximization problem.
\cite{ref_related_work_8_1} utilized the FA to suppress the self-interference in a monostatic full-duplex (FD) ISAC system,
and further promote the weighted sum of communication rate and sensing MI.
In the FA-enabled FD ISAC system,
the authors of \cite{ref_related_work_9} investigated the transmit power consumption minimization problem 
while considering discrete candidate positions for FA.
Besides, 
in \cite{ref_related_work_9_1},
the authors considered maximizing the weighted sum of sensing and communication rates (WSR) in an FA-aided near-field FD ISAC system
and proposed an antenna position matching (APM) algorithm for reducing the antenna movement distance.
The work \cite{ref_related_work_10} demonstrated that the FA could remarkably enhance the communication capacity for
both DL and uplink (UL) users in the networked FD ISAC system.
The paper \cite{ref_related_work_new_4}  incorporated the FA into an MIMO ISAC system 
to maximize the signal-clutter-noise ratio (SCNR) while satisfying communication SINR requirements.
The work \cite{ref_related_work_new_4_1} investigated the DL and UL sum-rate maximization problem in a FAS-assisted FD ISAC system.
Lately, 
a novel optimization algorithm based on a two-timescale framework for FA-enabled ISAC systems, 
tackling key issues such as slow antenna movement speed, dynamic RCS variation, 
and imperfect channel state information (CSI), 
was developed in \cite{ref_related_work_new_5}.

\subsection{Motivations and Contributions}
As seen above,
although a substantial amount of research has explored joint beamforming and position optimization for FA-assisted ISAC systems,
most of these works \cite{ref_related_work_new_1}$-$\cite{ref_related_work_8}, \cite{ref_related_work_new_4}$-$\cite{ref_related_work_new_5} have focused on integrated sensing and \textit{downlink} communication systems.
Research on FA-enabled integrated sensing and \textit{uplink} communication systems remains limited and is still emerging.
Besides, recent investigations \cite{ref_related_work_8_1}$-$\cite{ref_related_work_10} 
have considered the FA-aided FD ISAC system containing DL and UL communication. 
However, 
the sensing performance metrics used in \cite{ref_related_work_8_1}$-$\cite{ref_related_work_10},
which include MI \cite{ref_related_work_8_1} and the sensing SINR \cite{ref_related_work_9}$-$\cite{ref_related_work_10},
are challenging to quantify explicitly.
In contrast, 
the well-known CRB is a widely accepted metric for estimating sensing performance, 
providing a lower bound on the variance of unbiased parameter estimators with closed-form expressions.
Motivated by these gaps, 
our study focuses on an FA-aided uplink ISAC system aiming to enhance both its communication and estimation capabilities.
The contributions of this paper are detailed as follows:

\begin{itemize}
\item
This paper investigates the joint optimization of beamforming and the positions of FAs in a UL ISAC system enhanced by FA technology to boost communication and estimation performances.
Our objective is to minimize the CRB for estimating the target's DoA while simultaneously assuring the sum-rate of all UL users via designing BS probing beamforming, UL users' power allocation, receiving filters and FAs' position coefficients.
To the best of our knowledge, this problem has rarely been considered in the existing literature, e.g., \cite{ref_related_work_new_1}$-$\cite{ref_related_work_new_5}.

\item
By introducing splitting variables and leveraging the penalty dual decomposition (PDD) \cite{ref_PDD_1} framework combined with the majorization-minimization (MM) \cite{ref_MM} method, 
we develop a low-complexity algorithm to address the challenging 
beamformer configuration problem containing  \textit{fractional} and \textit{quartic} terms.
To the best of our knowledge, 
this has not been studied  in the existing literature \cite{ref_related_work_new_1}$-$\cite{ref_related_work_new_5}.
Besides, we also derive the closed-form solution for the FAs' position coefficients.

\item
Furthermore,
we propose an iteration optimization algorithm that analytically optimizes all variables using convex optimization techniques to effectively tackle the non-convex CRB minimization problem, 
without relying on any numerical solvers such as CVX \cite{ref_CVX}. 
This approach is notably rare in the existing literature, 
e.g., \cite{ref_related_work_new_1}$-$\cite{ref_related_work_new_5}.

\item
Extensive numerical results are presented to demonstrate the significant benefits of employing FAs to enhance estimation accuracy in the UL ISAC system.
Additionally,
our proposed PDD-based algorithm (i.e., Alg. \ref{alg:PDD}) 
showcases the superior efficiency compared to the method presented in \cite{ref_Convex Optimization} (i.e., the Schur complement).

\end{itemize}

The rest of the paper is organized as follows. 
Section II will introduce the model of the UL ISAC system assisted by FA
and formulate the joint beamforming and position coefficients design problem. 
Section III will propose a low-complexity iterative solution to tackle the proposed problem. 
Section IV and Section V will present numerical results and conclusions of the paper, respectively.

\section{System Model and Problem Formulation}
\subsection{System Model}

\begin{figure}[t]
	\centering
	\includegraphics[width=.45\textwidth]{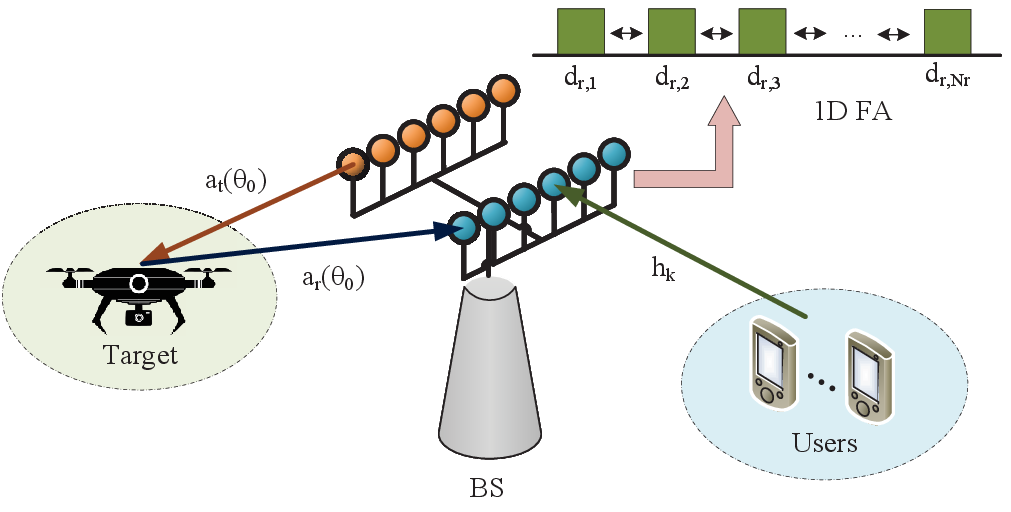}
	\caption{An FA-aided uplink ISAC system.}
	\label{fig.1}
\end{figure}

As shown in Fig. \ref{fig.1},
we consider an FA-aided UL communication ISAC system
comprising one base station (BS) equipped with uniform linear array (ULA) with $N_t$ transmit FPAs and $N_r$ receive FAs,
$K$ single-antenna uplink mobile users,
and one point-like sensing target 
\footnote{
Future work will consider an FA-aided ISAC system containing RCS fluctuations, multipath, and clutter, as well as accounting for propagation delays and Doppler effects.
}
In this system, 
BS sends probing waveform to estimate the target's angle parameter
and receives the information from UL users and echo signal from the target simultaneously.
During the whole procedure, 
all mobile users operate in uplink mode and transmit information symbols to the BS. 
For convenience, 
the sets of users, 
BS' transmit antennas and receive antennas 
are denoted as 
$\mathcal{K}$,
$\mathcal{N}_t$
and
$\mathcal{N}_r$,
respectively.

Furthermore,
the FAs, which are based on mechanically \cite{ref_channel model_1}$-$\cite{ref_channel model_2} or liquid-based \cite{FAS_3}$-$\cite{FAS_4} elements,
can move within a local region in real time.
The antenna positioning vectors (APVs) of BS transmit antennas and receive antennas are given by
$\mathbf{d}_t \triangleq [ d_{t,1}, d_{t,2}, \cdots , d_{t,N_t} ]^T $
and
$\mathbf{d}_r \triangleq [ d_{r,1}, d_{r,2}, \cdots , d_{r,N_r} ]^T $
within the given line segment of length $d_{max}$,
respectively.

According to the far-field wireless channel model \cite{ref_channel model_1}$-$\cite{ref_channel model_2},
changing the positions of FAs will influence the complex path coefficient,
while does not affect the angle of departure (AoD), 
the angle of arrival (AoA), 
and the amplitude of the complex path coefficient.
Moreover,
we assume that the channel is quasi-static \cite{ref_channel model_1}$-$\cite{ref_channel model_2}.
Let $L_{t,k}$ and $L_{r,k}$ denote the total number of transmit 
and receive channel paths at the BS from the $k$-th UL user, respectively. 
For the $i$-th transmit path between BS  and user $k$,
the  AoD  is
given as 
$\theta^t_{k,i} \in [-\frac{\pi}{2}, \frac{\pi}{2}] $.
For the $i$-th receive path between BS  and user $k$,
the elevation and azimuth AoAs are represented as 
$\theta^r_{k,i} \in [-\frac{\pi}{2}, \frac{\pi}{2}] $
and 
$\phi^r_{k,i} \in [-\frac{\pi}{2}, \frac{\pi}{2}] $,
respectively.

The transmit field response matrix of all $N_r$ FAs from the BS to the $k$-th user is represented as 
\begin{align}
\mathbf{H}_{0,k}(\mathbf{d}_{r}) = [\mathbf{h}_{1,k,1},\mathbf{h}_{1,k,2}, \cdots, \mathbf{h}_{1,k,N_t} ] \in \mathbb{C}^{L_{t,k} \times N_r },
\end{align}
where the field-response vector is 
$\mathbf{h}_{1,k,n}(d_{r,n})\! \triangleq \! [e^{j\frac{2\pi}{\lambda}d_{r,n}\sin(\theta^t_{k,1})},
e^{j\frac{2\pi}{\lambda}d_{r,n}\sin(\theta^t_{k,2})},
\cdots,
e^{j\frac{2\pi}{\lambda}d_{r,n}\sin(\theta^t_{k,L_{t,k}}) } ]^T \in \mathbb{C}^{L_{t,k} \times 1 } $,
and $ \lambda$ is the wavelength.
$\mathbf{h}_{0,k}\in \mathbb{C}^{ L_{r,k} \times 1}$ is defined as 
the receive field-response vector  from the BS to the $k$-th user.
Therefore, 
the channel vector between the BS and the $k$-th UL user is given by
\begin{align}
\mathbf{h}_k^H
=
\mathbf{h}_{0,k}^H
\bm{\Sigma}_k
\mathbf{H}_{0,k}(\mathbf{d}_{r})\in \mathbb{C}^{1 \times N_r },
\end{align}
where
$\bm{\Sigma}_k\in \mathbb{C}^{L_{r,k} \times L_{t,k} }$
is  denoted as the response of
all transmit and receive paths from the BS to 
the $k$-th UL user.

Besides,
 the transmit and receive steering vectors of the BS towards the sensing target  
 are respectively represented by
\begin{align}
&\mathbf{a}_{t,1}\! \!\triangleq \!\! [e^{j\frac{2\pi}{\lambda}\!d_{t,1}\!\sin(\!\theta_{0}\!)},
e^{j\frac{2\pi}{\lambda}\!d_{t,2}\!\sin(\!\theta_{0}\!)},
\cdots,
e^{j\frac{2\pi}{\lambda}\!d_{t,N_t}\!\sin(\!\theta_{0}\!) } ]^T,\\
&\mathbf{a}_{r,1}\! \!\triangleq \!\! [e^{j\frac{2\pi}{\lambda}\!d_{r,1}\!\sin(\!\theta_{0}\!)},
e^{j\frac{2\pi}{\lambda}\!d_{r,2}\!\sin(\!\theta_{0}\!)},
\cdots,
e^{j\frac{2\pi}{\lambda}\!d_{r,N_r}\!\sin(\!\theta_{0}\!) } ]^T,
\end{align}
where 
$\theta_{0}$ 
denotes the target's DoA.
Furthermore, 
the complex path coefficients of the transmit and receive steering vectors 
are denoted as 
$\beta_t$ 
and  
$\beta_r$,
respectively.
Then,
the transmit and receive line-of-sight (LoS) channels are respectively given as
\begin{align}
&\mathbf{a}_{t}(\theta_{0}) = \beta_t\mathbf{a}_{t,1}(\theta_{0}),
\mathbf{a}_{r}(\theta_{0}) = \beta_r\mathbf{a}_{r,1}(\theta_{0}),
\end{align}

The uplink signal transmitted at time slot $l$ of the $k$-th UL user is represented as
\begin{align}
x^u_k[l] = \sqrt{q}_ks^u_k[l], \forall k \in \mathcal{K},
\end{align}
with
$s^u_k[l]$ 
and
${q}_k$
representing
the information symbol and transmission power of the $k$-th UL user, respectively.
Besides,
$s^u_k[l]$ 
are assumed as mutually uncorrelated and
each has zero mean and unit variance.

Let 
$\mathbf{W} \in \mathbb{C}^{N_t \times N_t} $ 
denote the beamforming matrix for radar sensing.
The probing signal transmitted in the $l$-th time slot is given by
\begin{align}
\mathbf{x}[l] = \mathbf{W}\mathbf{s}_r[l],
\end{align}
where
$\mathbf{s}_r[l] \in \mathbb{C}^{N_t \times 1} $
denotes radar probing signal
and has zero mean and covariance matrix 
$ \mathbb{E}\{\mathbf{s}_r[l]\mathbf{s}^H_r[l]\}= \mathbf{I}_{N_t} $.

The received signal at the BS at time slot $l$ is written as
\begin{align}
\mathbf{y}[l] = {\sum}_{k=1}^{K} \mathbf{h}_kx^u_k[l]
+ \alpha\mathbf{a}_{r}\mathbf{a}_{t}^H\mathbf{x}[l] + \mathbf{n}[l],\label{received signal at the BS}
\end{align}
where 
$\alpha \in \mathbb{C} $ 
denotes the RCS coefficient
and
$\mathbf{n}[l] \sim \mathcal{CN}(0,\sigma^2\mathbf{I}_{N_t})$
denotes the additive white Gaussian
noise (AWGN) at the BS receiver.

To decode UL users' information $\{s^u_k[l]\}$,
the BS adopts the linear filter $\mathbf{u}_k \in \mathbb{C}^{N_r \times 1}$
to post-process the received signal.
Therefore, the output of the $k$-th filter can be given as
\begin{align}
{y}_k[l] \!=\! \mathbf{u}_k^H({\sum}_{k=1}^{K} \mathbf{h}_kx^u_k[l]
\!+\! \alpha\mathbf{a}_{r}\mathbf{a}_{t}^H\mathbf{x}[l] + \mathbf{n}[l]),
\forall k \in \mathcal{K}.
\end{align}

Then, 
the SINR of UL user $k$ can be obtained as
\begin{align}
\text{SINR}_k = \frac{{q}_k\vert\mathbf{u}_k^H\mathbf{h}_k\vert^2}
{\sum_{i \neq k}^{K} {q}_i\vert\mathbf{u}_k^H\mathbf{h}_i\vert^2 
+ \Vert \mathbf{u}_k^H \alpha\mathbf{a}_{r}\mathbf{a}_{t}^H\mathbf{W}\Vert^2_2
 + \sigma^2\Vert\mathbf{u}_k^H\Vert^2_2 },
\end{align}
and
the achievable rate of $k$-th user is given as
\begin{align}
\text{R}_k = \text{log}(1+\text{SINR}_k).
\end{align}

After BS recovers the communication information at the BS receiver,
it will adopt the successive interference cancellation (SIC) technique \cite{ref_SIC_1}$-$\cite{ref_SIC_3} 
to eliminate\footnote{
This work adopts the perfect SIC assumption as an upper performance bound.
The case of imperfect SIC will be studied in future work.}
the uplink communication signals from the received signals (\ref{received signal at the BS}).
Therefore,
the signal for estimating target's location information at the time slot $l$ is written as
\begin{align}
\mathbf{y}_r[l] =  \alpha \mathbf{A}(\theta_{0})\mathbf{x}[l] + {\mathbf{n}}[l].
\end{align}
where 
$\mathbf{A}(\theta_{0}) \triangleq \mathbf{a}_{r}(\theta_{0})\mathbf{a}_{t}^H(\theta_{0}) $.
By stacking $L$ coherent time slots,
the received echo signals can be given as 
\begin{align}
\mathbf{Y}_r = \alpha\mathbf{A}(\theta_{0})\mathbf{W}\mathbf{S}_r + \mathbf{{N}}, \label{YY_r}
\end{align}
where 
$ \mathbf{S}_r \triangleq [ \mathbf{s}_r[1], \mathbf{s}_r[2], \cdots ,\mathbf{s}_r[L] ] $
and
$  \mathbf{{N}} \triangleq [{\mathbf{n}}[1], {\mathbf{n}}[2], \cdots , {\mathbf{n}}[L]] $.
Moreover,
when $L$ is sufficiently large, 
the sample covariance matrix of $\mathbf{S}_r$ can be approximated by
\begin{align}
\frac{1}{L}\mathbf{S}_r\mathbf{S}_r^H  \thickapprox \mathbf{I}.
\end{align}

Note that CRB is a  lower bound of any unbiased estimator for describing the parameter estimation accuracy.
In the following, 
we will derive the closed-form CRB expression for target's DoA $\theta_{0}$.
Firstly,
we define the target parameters as
$\bm{\zeta} \triangleq [\theta_{0}, \alpha_R, \alpha_I]^T $,
where
$\alpha_R = \Re\{\alpha\}$
$\alpha_I = \Im\{\alpha\}$
and
$\bm{\tilde{\alpha}} \triangleq [\alpha_R, \alpha_I]^T $.

According to \cite{ref_CRB},
the Fisher information matrix (FIM) for the estimation of 
$\bm{\zeta}$ 
is given by
\begin{align}
\mathbf{F} = \mathbb{E}\bigg[ \frac{\partial \mathrm{In}(f(\mathbf{Y}_r\vert\bm{\zeta} )) }
{\partial \bm{\zeta} }\bigg(  \frac{\partial \mathrm{In}(f(\mathbf{Y}_r\vert\bm{\zeta} )) }
{\partial \bm{\zeta} } \bigg)^H  \bigg],\label{FIM_O}
\end{align}
where 
$f(\mathbf{Y}_r\vert\bm{\zeta} )$
denotes the conditional probability density function (PDF)
of $\mathbf{Y}_r$ given $\bm{\zeta}$

Next,
we will show the  explicit expression of
the FIM in (\ref{FIM_O}).
The joint conditional distribution of 
$\mathbf{Y}_r$ given
$\bm{\zeta}$
 can be written as
\begin{align}
f(\mathbf{Y}_r\vert\bm{\zeta} )
=
\frac{1}{\pi^{N_rL}\vert\sigma^2\mathbf{I}_{N_rL}\vert}
\text{exp}\bigg\{ -\frac{\Vert\mathbf{Y}_r - \alpha\mathbf{A}(\theta_{0})\mathbf{W}\mathbf{S}_r \Vert^2_F}
{\sigma^2} \bigg\}.
\end{align}
Therefore,
the log-likelihood function for estimating
$\bm{\zeta}$
based on the observation 
$\mathbf{Y}_r$ 
is given by
\begin{align}
&\mathrm{In}(f(\mathbf{Y}_r\vert\bm{\zeta} ))
=
-N_rL\mathrm{In}(\pi\sigma^2)
-\frac{\Vert\mathbf{Y}_r \Vert^2_F}{\sigma^2}\label{likelihood_function}\\
& - \frac{\vert\alpha\vert^2
\Vert\mathbf{A}(\theta_{0})\mathbf{W}\mathbf{S}_r \Vert^2_F}{\sigma^2}
+ \frac{2\Re\{ \mathrm{Tr}( \alpha^{\ast} \mathbf{S}_r^H\mathbf{W}^H\mathbf{A}(\theta_{0})^H\mathbf{Y}_r  ) \}}{\sigma^2},\nonumber
\end{align}

According to \cite{ref_FIM_1}
and based on 
(\ref{FIM_O})
and
(\ref{likelihood_function}),
$\mathbf{F}$
can be directly given as
\begin{align}
\mathbf{F}
=
\begin{bmatrix}
\mathrm{F}_{\theta_{0}\theta_{0}} \ & \mathbf{F}_{\theta_{0}\alpha}\\
\mathbf{F}_{\theta_{0}\alpha}^H \ & \mathbf{F}_{\alpha\alpha}
\end{bmatrix}\in \mathbb{C}^{3\times 3}.\label{F_O_expression}
\end{align}
Besides,
the elements of $\mathbf{F}$ are given as
\begin{align}
&\mathrm{F}_{\theta_{0}\theta_{0}} 
=
\frac{2L\vert\alpha\vert^2}{\sigma^2} \mathrm{tr}(\mathbf{A}_1 \mathbf{W}\mathbf{W}^H),\\
&\mathbf{F}_{\theta_{0}\alpha}
=
\frac{2L}{\sigma^2}\Re\{ \alpha^{\ast} 
\mathrm{tr}(\mathbf{A}_2 \mathbf{W}\mathbf{W}^H)[1,j] \}  \\
&\mathbf{F}_{\alpha\alpha}
=
\frac{2L}{\sigma^2} \mathrm{tr}(\mathbf{A}_3 \mathbf{W}\mathbf{W}^H)\mathbf{I}_2,
\end{align}
respectively,
where the newly introduced coefficients are defined 
in (\ref{AA_1}),
respectively,
with
$\dot{\mathbf{a}}(\theta_{0})$ denoting the derivative of $\mathbf{a}(\theta_{0})$ with respect to (w.r.t.) $\theta_{0}$.
The details of the derivation of the FIM  $\mathbf{F}$ 
can be seen in Appendix A.

\begin{figure*}
\begin{align}
&\mathbf{A}_1 \triangleq
{\sum}_{n=1}^{N_r}
(\frac{2\pi}{\lambda}d_{r,n}\cos(\theta_{0}))^2
\mathbf{a}_{t}(\theta_{0})\mathbf{a}_{t}^H(\theta_{0})
+
({\sum}_{n=1}^{N_r} -j\frac{2\pi}{\lambda} d_{r,n}\cos(\theta_{0})
\mathbf{a}_{t}(\theta_{0})
\dot{\mathbf{a}}_{t}^H(\theta_{0})
 )\label{AA_1}\\
 &+
 ({\sum}_{n=1}^{N_r} j\frac{2\pi}{\lambda} d_{r,n}\cos(\theta_{0})
\dot{\mathbf{a}}_{t}(\theta_{0})
\mathbf{a}_{t}^H(\theta_{0})
 )
 +
 N_r\dot{\mathbf{a}}_{t}(\theta_{0})
 \dot{\mathbf{a}}_{t}^H(\theta_{0})
,\nonumber
 \\
&\mathbf{A}_2 \triangleq
({\sum}_{n=1}^{N_r} -j\frac{2\pi}{\lambda}d_{r,n}\cos(\theta_{0}) )
\mathbf{a}_{t}(\theta_{0})\mathbf{a}_{t}^H(\theta_{0})
+ N_r  \dot{\mathbf{a}}_{t}(\theta_{0})\mathbf{a}_{t}^H(\theta_{0}),
\mathbf{A}_3 \triangleq
 N_r {\mathbf{a}}_{t}(\theta_{0})\mathbf{a}_{t}^H(\theta_{0}).\nonumber
\end{align}
\boldsymbol{\hrule}
\end{figure*}

Next,
the CRB for estimating the angle 
$\theta_0$,
which corresponds to the first diagonal unit of 
$\mathbf{F}^{-1}$,
can be given as
\begin{align}
&\mathrm{CRB}_{\theta_0}
=
[\mathbf{F}^{-1}]_{1,1}\\
&=
\frac{\sigma^2}{2L \vert\alpha\vert^2 \bigg( 
\mathrm{tr}(\mathbf{A}_1 \mathbf{W}\mathbf{W}^H)
-
\frac{\vert\mathrm{tr}(\mathbf{A}_2 \mathbf{W}\mathbf{W}^H)\vert^2}{\mathrm{tr}(\mathbf{A}_3 \mathbf{W}\mathbf{W}^H)} 
\bigg)}.\nonumber
\end{align}

\subsection{Problem Formulation}

Our objective is to minimize the CRB of estimating the angle information $\theta_0$
via jointly optimizing 
the transmit beamformer $\mathbf{W}$,
the linear filters $\{\mathbf{u}_k\}$,
the UL users' transmit power $\{q_k\}$
and the FA position coefficients $\{\mathbf{d}_r\}$.
The optimization problem 
\footnote{{The two-time scale optimization problem will be investigated in future work.}}
can be mathematically formulated as
\begin{subequations}
\begin{align}
\textrm{(P0)}:
&\mathop{\textrm{min}}
\limits_{
\mathbf{W},
\{\mathbf{u}_k\},
\{q_k\},
\mathbf{d}_r
}\
\mathrm{CRB}_{\theta_0}\\
\textrm{s.t.} &
{\sum}_{k=1}^{K}\text{R}_k \geq {R}_{t},\label{P0_c1}\\
& \Vert\mathbf{W}\Vert_F^2 \leq P_{BS},\\
& 0 \leq q_k \leq P_{u,k}, \forall k \in \mathcal{K},\\
& d_{r,1} \geq 0,
d_{r,N_r} \leq d_{max},\\
&d_{r,n}-d_{r,n-1} \geq d_{min},
n = 2,3, \cdots, N_r,
\label{P0_c6}
\end{align}
\end{subequations}
where
${R}_{t}$ is the predefined sum-rate threshold of all UL users,
$P_{BS}$
and
$P_{u,k}$ denote the  maximum transmission power of the
BS and the $k$-th UL user,
respectively,
$d_{max}$ represents the maximum moving distance of BS's FA
and 
$d_{min}$ is the minimum distance between adjacent antennas for avoiding
antenna coupling effect.

\section{Algorithm Design}
\subsection{Problem Reformulation}

Firstly,
since
minimizing
$\mathrm{CRB}_{\theta_0}$
is equivalent to maximizing
$\mathrm{tr}(\mathbf{A}_1 \mathbf{W}\mathbf{W}^H)
-
\frac{\vert\mathrm{tr}(\mathbf{A}_2 \mathbf{W}\mathbf{W}^H)\vert^2}{\mathrm{tr}(\mathbf{A}_3 \mathbf{W}\mathbf{W}^H)}$,
we turn to solve the following problem:
\begin{subequations}
\begin{align}
\textrm{(P1)}:
\mathop{\textrm{max}}
\limits_{
\mathbf{W},
\{\mathbf{u}_k\},
\atop
\{q_k\},
\mathbf{d}_r
}\
&
\mathrm{tr}(\mathbf{A}_1 \mathbf{W}\mathbf{W}^H)
-
\frac{\vert\mathrm{tr}(\mathbf{A}_2 \mathbf{W}\mathbf{W}^H)\vert^2}{\mathrm{tr}(\mathbf{A}_3 \mathbf{W}\mathbf{W}^H)}\\
\textrm{s.t.} &\
(\mathrm{\ref{P0_c1}})-(\mathrm{\ref{P0_c6}}).\nonumber
\end{align}
\end{subequations}

Besides,
to make the problem (P1) more tractable,
we will use the fractional programming (FP) method \cite{ref_FP}
to equivalently transform the sum-rate constraint (\ref{P0_c1}).
Firstly,
by leveraging the Lagrangian dual reformulation
 and
introducing
auxiliary variables 
$\bm{\gamma} = [\gamma_1, \cdots, \gamma_K]^T$, 
the original sum-rate constraint (\ref{P0_c1})
 can be written in (\ref{FP_1}).
\begin{figure*}
\begin{align}
\textrm{R}_1
=
{\sum}_{k=1}^{K}
\bigg(
\textrm{log}(1+\gamma_k)
-
\gamma_k
+
\frac{(1+\gamma_k){q}_k\vert\mathbf{u}_k^H\mathbf{h}_k\vert^2}
{
\sum_{i =1}^{K} {q}_i\vert\mathbf{u}_k^H\mathbf{h}_i\vert^2 
+ \Vert \mathbf{u}_k^H \alpha\mathbf{a}_{r}\mathbf{a}_{t}^H\mathbf{W}\Vert^2_2
 + \sigma^2\Vert\mathbf{u}_k^H\Vert^2_2
}
\bigg)\label{FP_1}
\end{align}
\bm{\hrule}
\end{figure*}
And then, 
by exploiting the quadratic transform
and
introducing the auxiliary variables
$\bm{\omega} = [\omega_1,  \cdots, \omega_{K}  ]^T$,
equation (\ref{FP_1})
can be further converted to (\ref{FP_2}).
\begin{figure*}
\begin{align}
\textrm{R}_2
=
{\sum}_{k=1}^{K}
\bigg(
\textrm{log}(1+\gamma_k)
-
\gamma_k
+2\sqrt{(1+\gamma_k)}\Re\{\omega_k^{\ast} \sqrt{{q}_k} \mathbf{u}_k^H\mathbf{h}_k \}  
-\vert\omega_k\vert^2
( 
{\sum}_{i =1}^{K} {q}_i\vert\mathbf{u}_k^H\mathbf{h}_i\vert^2 
+ \Vert \mathbf{u}_k^H \alpha\mathbf{a}_{r}\mathbf{a}_{t}^H\mathbf{W}\Vert^2_2
 + \sigma^2\Vert\mathbf{u}_k^H\Vert^2_2
)
\bigg)\label{FP_2}
\end{align}
\bm{\hrule}
\end{figure*}

Based on the above transformation,
the problem (P1) can be reexpressed as
\begin{subequations}
\begin{align}
\textrm{(P2)}:
\mathop{\textrm{max}}
\limits_{
\mathbf{W},
\{\mathbf{u}_k\},
\atop
\{q_k\},
\mathbf{d}_r
}\
&
\mathrm{tr}(\mathbf{A}_1 \mathbf{W}\mathbf{W}^H)
-
\frac{\vert\mathrm{tr}(\mathbf{A}_2 \mathbf{W}\mathbf{W}^H)\vert^2}{\mathrm{tr}(\mathbf{A}_3 \mathbf{W}\mathbf{W}^H)}\\
\textrm{s.t.}\ &
\text{R}_2 \geq {R}_{t},\label{P2_c1}\\
& \Vert\mathbf{W}\Vert_F^2 \leq P_{BS},\\
& 0 \leq q_k \leq P_{u,k}, \forall k \in \mathcal{K},\\
& d_{r,1} \geq 0,
d_{r,N_r} \leq d_{max},\\
&d_{r,n}\!\!-\!d_{r,n\!-\!1}\!\! \geq\!\! d_{min},
n\!\! =\!\! 2,3, \cdots, N_r.\label{P2_c6}
\end{align}
\end{subequations}

In the next, 
we propose the block coordinate ascent (BCA)-based algorithm
\cite{ref_BCA}
 to tackle the problem (P2).

\subsection{Optimizing Auxiliary Variables}
According to the derivation of FP method \cite{ref_FP},
the auxiliary variables 
$\bm{\gamma}$
and
$\bm{\omega}$
can be updated by analytical solutions
that are respectively given as follows
\begin{align}
&\gamma_k^{\star}
\!=\!
\frac{{q}_k\vert\mathbf{u}_k^H\mathbf{h}_k\vert^2}
{{\sum_{i \neq k }^{K} {q}_i\vert\mathbf{u}_k^H\mathbf{h}_i\vert^2 
\!+\! \Vert \mathbf{u}_k^H \alpha\mathbf{a}_{r}\mathbf{a}_{t}^H\mathbf{W}\Vert^2_2
 \!+\! \sigma^2\Vert\mathbf{u}_k^H\Vert^2_2 }},\label{FP_variable_1}
\\
&\omega_k^{\star}
\!=\!
\frac{\sqrt{1+\gamma_k}( \sqrt{{q}_k}\mathbf{u}_k^H\mathbf{h}_k) }
{
\sum_{i =1}^{K} {q}_i\vert\mathbf{u}_k^H\mathbf{h}_i\vert^2 
\!+\! \Vert \mathbf{u}_k^H \alpha\mathbf{a}_{r}\mathbf{a}_{t}^H\mathbf{W}\Vert^2_2
\! +\! \sigma^2\Vert\mathbf{u}_k^H\Vert^2_2 
}.\label{FP_variable_2}
\end{align}

\subsection{Updating The BS Beamformer}
In this subsection,
when other variables are given,
we investigate the update of the BS beamformer $\mathbf{W}$.
By
defining the new coefficients as follows
\begin{align}
&\mathbf{B}_{2}
\triangleq
{\sum}_{k=1}^{K}\big(
\mathbf{I}_{N_t\cdot N_t} \otimes 
(\vert\alpha\vert^2 \mathbf{a}_{t}\mathbf{a}_{r}^H\mathbf{u}_k
\mathbf{u}_k^H\mathbf{a}_{r}\mathbf{a}_{t}^H
 )\big),\\
&c_1
\triangleq
{\sum}_{k=1}^{K}
\bigg(
\textrm{log}(1+\gamma_k)
\!-\!
\gamma_k
\!+\!
2\sqrt{(1\!\!+\!\!\gamma_k)}\Re\{\omega_k^{\ast}\sqrt{{q}_k} \mathbf{u}_k^H\mathbf{h}_k \}  
\nonumber\\
&-\vert\omega_k\vert^2
( 
{\sum}_{i =1}^{K} {q}_i\vert\mathbf{u}_k^H\mathbf{h}_i\vert^2 
 + \sigma^2\Vert\mathbf{u}_k^H\Vert^2_2
)
\bigg)\!-\!R_t,
\mathbf{w}
\triangleq
\mathrm{vec}(\mathbf{W}),
\nonumber\\
 &\mathbf{B}_{3}
 \triangleq
 \mathbf{I}_{N_t\cdot N_t} \otimes \mathbf{A}_1,
 \mathbf{B}_{4}
 \triangleq
 \mathbf{I}_{N_t\cdot N_t} \otimes \mathbf{A}_2,
 \mathbf{B}_{5}
 \triangleq
 \mathbf{I}_{N_t\cdot N_t} \otimes \mathbf{A}_3.\nonumber
\end{align}

Based on the above transformation, 
the optimization problem w.r.t. $\mathbf{w}$
can be rewritten as
\begin{subequations}
\begin{align}
\textrm{(P3)}:
\mathop{\textrm{min}}
\limits_{
\mathbf{w}
}\
&
\frac{(\mathbf{w}^H \mathbf{B}_{4} \mathbf{w} )
(\mathbf{w}^H \mathbf{B}_{4}^H \mathbf{w} )}
{\mathbf{w}^H \mathbf{B}_{5} \mathbf{w} }-
\mathbf{w}^H \mathbf{B}_{3} \mathbf{w} \label{P3_obj}
\\
\textrm{s.t.}\ &
\mathbf{w}^H \mathbf{B}_{2} \mathbf{w} \leq c_1,\label{P3_c1}\\
& \mathbf{w}^H \mathbf{w} \leq P_{BS}.
\end{align}
\end{subequations}

Obviously,
the problem (P3) contains both the fractional  
and the quartic terms in the objective function (\ref{P3_obj})
is difficult to solve.

To address the challenge mentioned, 
we  introduce auxiliary variables $\mathbf{f}$
and $b$ to decouple the objective function (\ref{P3_obj}).
Thus,
the problem (P3) can be equivalently written as
\begin{subequations}
\begin{align}
\textrm{(P4)}:
\mathop{\textrm{min}}
\limits_{
\mathbf{w},
\mathbf{f},
b
}\
&
\frac{(\mathbf{w}^H \mathbf{B}_{4} \mathbf{f} )
(\mathbf{f}^H \mathbf{B}_{4}^H \mathbf{w} )}
{b }-
\mathbf{w}^H \mathbf{B}_{3} \mathbf{w} \label{P4_obj}
\\
\textrm{s.t.}\ &
\mathbf{w}^H \mathbf{B}_{2} \mathbf{w} \leq c_1,\label{P4_c1}\\
& \mathbf{f}^H \mathbf{f} \leq P_{BS},\\
& \mathbf{w} = \mathbf{f},\label{P4_c3}\\
& \mathbf{w}^H \mathbf{B}_{5} \mathbf{f} = b\label{P4_c4}.
\end{align}
\end{subequations}

Next,
based on the PDD framework \cite{ref_PDD_1}, \cite{ref_PDD_2}$-$\cite{ref_PDD_3},
we propose a PDD-based efficient method to solve the problem (P4).
Firstly,
by penalizing 
the equality constraints
(\ref{P4_c3}) 
and
(\ref{P4_c4}) 
into the objective
function,
an
augmented Lagrangian (AL) minimization problem can be given as
\begin{subequations}
\begin{align}
\textrm{(P5)}:
\mathop{\textrm{min}}
\limits_{
\mathbf{w},
\mathbf{f},
b
}\
&
\frac{(\mathbf{w}^H \mathbf{B}_{4} \mathbf{f} )
(\mathbf{f}^H \mathbf{B}_{4}^H \mathbf{w} )}
{b }-
\mathbf{w}^H \mathbf{B}_{3} \mathbf{w} \label{P5_obj}\\
&+ \frac{1}{2\rho}\Vert\mathbf{w} - \mathbf{f}\Vert^2_2
 + \Re\{ \bm{\lambda}_1^H(\mathbf{w} - \mathbf{f}) \}\nonumber \\
&+\frac{1}{2\rho}\vert \mathbf{w}^H \mathbf{B}_{5} \mathbf{f} - b\vert^2
+\Re\{ \lambda_2^{\ast}( \mathbf{w}^H \mathbf{B}_{5} \mathbf{f} - b ) \}\nonumber
\\
\textrm{s.t.}\ &
\mathbf{w}^H \mathbf{B}_{2} \mathbf{w} \leq c_1,\label{P5_c1}\\
& \mathbf{f}^H \mathbf{f} \leq P_{BS}.
\end{align}
\end{subequations}

Following the PDD method 
\cite{ref_PDD_1}, \cite{ref_PDD_2}$-$\cite{ref_PDD_3},
we conduct a two-layer iterative process,
which its inner layer updates
$\mathbf{w}$,
$\mathbf{f}$
and
$b$
by using a block coordinate descent (BCD) method
and its outer
layer selectively updating the penalty coefficient $\rho$ 
or the dual variables $\{\bm{\lambda}_1, \lambda_2\}$.
The PDD procedure will be detailed in the subsequent discussion.

\emph{\textit{Inner Layer Procedure}}

For the inner layer iteration, 
we will sequentially update 
$\mathbf{w}$,
$\mathbf{f}$
and
$b$.
When 
$\mathbf{f}$
and
$b$
are fixed,
the AL minimization problem 
w.r.t.
$\mathbf{w}$
is reduced to as follows
\begin{subequations}
\begin{align}
\textrm{(P6)}:
\mathop{\textrm{min}}
\limits_{
\mathbf{w}
}\
&
\mathbf{w}^H \mathbf{B}_{6} \mathbf{w} 
-
2\Re\{ \mathbf{b}_1^H\mathbf{w} \}
+c_3
-
\mathbf{w}^H \mathbf{B}_{3} \mathbf{w} \label{P6_obj}\\
\textrm{s.t.}\ &
\mathbf{w}^H \mathbf{B}_{2} \mathbf{w} \leq c_1,\label{P6_c1}
\end{align}
\end{subequations}
where the above new coefficients are given as follows
\begin{align}
&\mathbf{B}_{6}
\triangleq
\mathbf{B}_{4} \mathbf{f} \mathbf{f}^H\mathbf{B}_{4}^H/b
+(\mathbf{I}+\mathbf{B}_{5} \mathbf{f} \mathbf{f}^H\mathbf{B}_{5}^H)/(2\rho),\\
&\mathbf{b}_1
\triangleq
(\mathbf{f}-\bm{\lambda}_1+b\mathbf{B}_5\mathbf{f})/(2\rho)
-\lambda_2^{\ast}\mathbf{B}_5\mathbf{f}/2,\nonumber\\
&c_3
\triangleq
(\Vert\mathbf{f}\Vert_2^2+\vert b\vert^2)/(2\rho)
-\Re\{\bm{\lambda}_1^H\mathbf{f}+\lambda_2^{\ast}b \}.\nonumber
\end{align}

Obviously,
the problem (P6) is difficult to solve due to the 
the non-convex objective function (\ref{P6_obj}).
Inspired by the MM method \cite{ref_MM}, \cite{ref_MM_2}$-$\cite{ref_MM_3},
we can establish a tight lower bound of the non-convex objective function (\ref{P6_obj}),
which is given as
\begin{align}
\mathbf{w}^H \mathbf{B}_{3} \mathbf{w}
&\geq
\mathbf{w}^H_0 \mathbf{B}_{3} \mathbf{w}_0
+2\Re\{ \mathbf{w}^H_0 \mathbf{B}_{3}(\mathbf{w}-\mathbf{w}_0) \},\label{WW_MM_1}\\
&=
2\Re\{ \mathbf{w}^H_0 \mathbf{B}_{3}\mathbf{w} \}
- (\mathbf{w}^H_0 \mathbf{B}_{3} \mathbf{w}_0)^{\ast}.\nonumber
\end{align}
where $\mathbf{w}_0$ 
is obtained from the last iteration.
Therefore,
the term $\mathbf{w}^H \mathbf{B}_{3} \mathbf{w}$
in the objective function (\ref{P6_obj})
can be replaced by (\ref{WW_MM_1}).
And then,
we turn to optimize a tight convex upper bound of the objective function (\ref{P6_obj}),
which is written as
\begin{subequations}
\begin{align}
\textrm{(P7)}:
\mathop{\textrm{min}}
\limits_{
\mathbf{w}
}\
&
\mathbf{w}^H \mathbf{B}_{6} \mathbf{w} 
-
2\Re\{ \mathbf{b}_2^H\mathbf{w} \}
+c_4
 \label{P7_obj}\\
\textrm{s.t.}\ &
\mathbf{w}^H \mathbf{B}_{2} \mathbf{w} \leq c_1,\label{P7_c1}
\end{align}
\end{subequations}
where
$\mathbf{b}_2 
\triangleq \mathbf{b}_1 +\mathbf{B}_3^H\mathbf{w}_0 $
and
$
c_4 
\triangleq
c_3 + (\mathbf{w}^H_0 \mathbf{B}_{3} \mathbf{w}_0)^{\ast}
$.
The problem (P7) is a typical second order cone program (SOCP)
and can be solved by numerical solvers,
i.e., CVX \cite{ref_CVX}.

However,
the above update method,
relying on numerical solvers, e.g., CVX,
has the following two main shortcomings:
\noindent
i) the complexity of solving SOCP problem via convex optimization solvers including CVX,
which adopts the interior point (IP) method \cite{ref_Convex Optimization},
will dramatically increase as the variables' dimension grows;

\noindent
ii)
the application of third-party solvers unavoidably 
raises the cost and complicates the implementation of the algorithm
due to licensing fees, the need for software installation and maintenance, 
and the platform requirements to support the solver.

Thus, 
we aim to develop an algorithm that ideally does not depend on CVX.
Firstly, 
to  solve the problem (P7) efficiently, 
we introduce the following lemma which has been proven in  
\cite{ref_Trust region}.

\begin{lemma} 
\label{lem:complex_Convex_trust_region problem}
Consider the following convex trust region problem:
\begin{subequations}
\begin{align}
(\textrm{P}_{Lm1}):\mathop{\textrm{min}}
\limits_{\mathbf{x}}\
& \mathbf{x}^H\mathbf{Q}\mathbf{x}-2\Re\{\mathbf{q}^H\mathbf{x}\}+q\\
\textrm{s.t.}\
&\Vert\mathbf{x}\Vert_2^2  \leq \bar{q}, \label{P_lemm_c1}
\end{align}
\end{subequations}
where 
$\mathbf{Q} \succcurlyeq 0$
and Slater's condition holds.
Besides,
the eigenvalue decomposition of 
$\mathbf{Q}$ 
is given as 
$\mathbf{Q} = \mathbf{U} \Lambda \mathbf{U}^H $.
Then the optimal solution to ($\textrm{P}_{Lm1}$) can be determined by
\begin{align}
\mathbf{x}^{\star} = \mathbf{U}(\Lambda+\nu^{\star}\mathbf{I})\mathbf{U}^H\mathbf{q},
\end{align}
where
the value of $\nu$  is non-negative 
and can be efficiently obtained by the Newton's method 
or bisection search.

\end{lemma}

To leverage Lemma 
\ref{lem:complex_Convex_trust_region problem},
the constraint (\ref{P7_c1}) should be written 
into the standard bounded norm constraint (\ref{P_lemm_c1}) 
 of 
($\textrm{P}_{Lm1}$).
Next,
according to the singularity of the matrix 
$\mathbf{B}_{2}$,
we need to consider the two different cases \cite{ref_Singularity},
which will be discussed later.

\textbf{\textit{\emph{CASE-I}}}:
\textbf{The matrix 
$\mathbf{B}_{2}$
is invertible}.
By introducing the following new notions
\begin{align}
\mathbf{\bar{w}}
\triangleq
\mathbf{B}_{2}^{\frac{1}{2}}\mathbf{w},
\mathbf{\bar{B}}_{6}
\triangleq
(\mathbf{B}_{2}^{-\frac{1}{2}})^H\mathbf{B}_{6}\mathbf{B}_{2}^{-\frac{1}{2}},
\mathbf{\bar{b}}_2
\triangleq
(\mathbf{B}_{2}^{-\frac{1}{2}})^H\mathbf{b}_2,\label{P7_case_1}
\end{align}
and the problem (P7)
can be equivalently rewritten as
\begin{subequations}
\begin{align}
\textrm{(P8)}:
\mathop{\textrm{min}}
\limits_{
\mathbf{\bar{w}}
}\
&
\mathbf{\bar{w}}^H \mathbf{\bar{B}}_{6} \mathbf{\bar{w}} 
-
2\Re\{ \mathbf{\bar{b}}_2^H\mathbf{\bar{w}} \}
+c_4
 \label{P8_obj}\\
\textrm{s.t.}\ &
\mathbf{\bar{w}}^H  \mathbf{\bar{w}} \leq c_1,\label{P8_c1}
\end{align}
\end{subequations}

Obviously, 
by  Lemma \ref{lem:complex_Convex_trust_region problem},
the analytical solution 
$\mathbf{\bar{w}}^{\star}$
can be efficiently obtained.
Therefore,
via (\ref{P7_case_1}),
the optimal solution 
$\mathbf{w}^{\star}$
can be given as
\begin{align}
\mathbf{w}^{\star} = \mathbf{B}_{2}^{-\frac{1}{2}}\mathbf{\bar{w}}^{\star}.
\end{align}

\textbf{\textit{\emph{CASE-II}}}:
\textbf{The matrix 
$\mathbf{B}_{2}$
is singular}.
To invoke Lemma \ref{lem:complex_Convex_trust_region problem},
by leveraging the MM methodology 
\footnote{ This method guarantees that the subproblems relying on 
$\mathbf{B}_{2}$
  satisfy the conditions of Lemma 1.},
we conduct a tight upper bound of the constraint (\ref{P7_c1}),
which is given as
\begin{align}
&\mathbf{w}^H \mathbf{B}_{2} \mathbf{w}-c_2
=
(\mathbf{w}-\mathbf{w}_0)^H 
\mathbf{B}_{2}
(\mathbf{w}-\mathbf{w}_0)\label{P7_case_2}\\
&+2\Re\{ (\mathbf{B}_{2}\mathbf{w})^H(\mathbf{w}-\mathbf{w}_0) \}
+\mathbf{w}_0^H 
\mathbf{B}_{2}
\mathbf{w}_0
-c_2\nonumber\\
&\leq
(\mathbf{w}-\mathbf{w}_0)^H 
(\mathbf{B}_{2}+\delta\mathbf{I})
(\mathbf{w}-\mathbf{w}_0)\nonumber\\
&+
2\Re\{ (\mathbf{B}_{2}\mathbf{w})^H(\mathbf{w}-\mathbf{w}_0) \}
+\mathbf{w}_0^H 
\mathbf{B}_{2}
\mathbf{w}_0
-c_2\nonumber\\
&=
\mathbf{w}^H 
(\mathbf{B}_{2}+\delta\mathbf{I})
\mathbf{w}
-2\Re\{ \mathbf{w}^H_0\delta\mathbf{I}\mathbf{w}  \}
+ \mathbf{w}^H_0(\delta\mathbf{I})\mathbf{w}_0-c_2\nonumber\\ 
&=
\mathbf{w}^H 
\mathbf{\hat{B}}_{2}
\mathbf{w}
-2\Re\{ \mathbf{\hat{b}}_{2}^H\mathbf{w}  \}
+ \hat{c}_2 ,\nonumber
\end{align}
where
$\delta$ is a small positive constant,
$\mathbf{\hat{B}}_{2}\triangleq\mathbf{B}_{2}+\delta\mathbf{I}$,
$\mathbf{\hat{b}}_{2}\triangleq  \delta\mathbf{w}_0$
and
$ \hat{c}_2\triangleq  \mathbf{w}^H_0(\delta\mathbf{I})\mathbf{w}_0-c_2$.

Based on the above MM transformation,
the constraint (\ref{P7_c1})
can be replaced by (\ref{P7_case_2}).
And then,
we turn to solve the following problem
\begin{subequations}
\begin{align}
\textrm{(P9)}:
\mathop{\textrm{min}}
\limits_{
\mathbf{w}
}\
&
\mathbf{w}^H \mathbf{B}_{6} \mathbf{w} 
-
2\Re\{ \mathbf{b}_2^H\mathbf{w} \}
+c_4
 \label{P9_obj}\\
\textrm{s.t.}\ &
\mathbf{w}^H 
\mathbf{\hat{B}}_{2}
\mathbf{w}
-2\Re\{ \mathbf{\hat{b}}_{2}^H\mathbf{w}  \}
+ \hat{c}_2 \leq 0,\label{P9_c1}
\end{align}
\end{subequations}

In order to adopt Lemma \ref{lem:complex_Convex_trust_region problem},
we need to transfer the constraint (\ref{P9_c1})
into the standard form of Lemma \ref{lem:complex_Convex_trust_region problem}.
Firstly,
we define the following new coefficients
\begin{align}
& \mathbf{\tilde{w}}
\triangleq
\mathbf{\hat{B}}_{2}^{\frac{1}{2}}\mathbf{w}
-
\mathbf{\hat{B}}_{2}^{-\frac{1}{2}}\mathbf{\hat{b}}_{2},
\mathbf{\tilde{B}}_{2} 
\triangleq 
\mathbf{\hat{B}}_{2}^{-\frac{1}{2}}\mathbf{B}_{6}
\mathbf{\hat{B}}_{2}^{-\frac{1}{2}},\label{P7_case_2_1}\\
&\mathbf{\tilde{b}}_2
\triangleq
\mathbf{\hat{B}}_{2}^{-\frac{1}{2}}\mathbf{{b}}_{2}
-
\mathbf{B}_{6}^H\mathbf{\hat{B}}_{2}^{-\frac{3}{2}}\mathbf{\hat{b}}_{2},\nonumber
\end{align}
and then,
the problem (P9)
can be equivalently written as
\begin{subequations}
\begin{align}
\textrm{(P10)}:
\mathop{\textrm{min}}
\limits_{
\mathbf{\tilde{w}}
}\
&
\mathbf{\tilde{w}}^H \mathbf{\tilde{B}}_{2}  \mathbf{\tilde{w}}
-
2\Re\{ \mathbf{\tilde{b}}_2^H\mathbf{\tilde{w}} \}
 \label{P10_obj}\\
\textrm{s.t.}\ &
\mathbf{\tilde{w}}^H 
\mathbf{\tilde{w}}
  \leq
  \mathbf{\hat{b}}_{2}^H
  \mathbf{\hat{B}}_{2}^{-1}
  \mathbf{\hat{b}}_{2} -\hat{c}_2.\label{P10_c1}
\end{align}
\end{subequations}

Up to here, 
 we can obtain analytically
the solution $\mathbf{\tilde{w}}^{\star}$ via invoking Lemma \ref{lem:complex_Convex_trust_region problem}.
Furthermore,
by (\ref{P7_case_2_1}),
the optimal solution 
$\mathbf{w}^{\star}$ is immediately given as
\begin{align}
\mathbf{w}^{\star} = 
\mathbf{\hat{B}}_{2}^{-\frac{1}{2}}
\mathbf{\tilde{w}}^{\star}
+
\mathbf{\hat{B}}_{2}^{-1}\mathbf{\hat{b}}_{2}.
\end{align}
 
When $\mathbf{w}$ and $b$ are given, 
the update of $\mathbf{f}$ 
is reduced to solving the following problem
\begin{subequations}
\begin{align}
\textrm{(P11)}:
\mathop{\textrm{min}}
\limits_{
\mathbf{f}
}\
&
\mathbf{f}^H \mathbf{B}_{7}  \mathbf{f}
-
2\Re\{ \mathbf{b}_2^H\mathbf{f} \}
+c_5
 \label{P11_obj}\\
\textrm{s.t.}\ &
\mathbf{f}^H 
\mathbf{f}
  \leq
  P_{BS},\label{P11_c1}
\end{align}
\end{subequations}
where the new coefficients are defined as
\begin{align}
&\mathbf{B}_{7} 
\triangleq
(\mathbf{B}_4^H\mathbf{w}\mathbf{w}^H\mathbf{B}_4)/b
+(\mathbf{I}+\mathbf{B}_5^H\mathbf{w}\mathbf{w}^H\mathbf{B}_5)/(2\rho),\\
&\mathbf{b}_3
\triangleq
(\mathbf{w}+b\mathbf{B}_5^H\mathbf{w})/(2\rho)
+
(\bm{\lambda}_1-{\lambda}_2\mathbf{B}_5^H\mathbf{w})/2,\nonumber\\
&
c_5
\triangleq
(\mathbf{w}^H\mathbf{w}+\vert b\vert^2)/(2\rho)
+\Re\{\bm{\lambda}_1^H\mathbf{w} - \lambda_2^{\ast}b\}-
\mathbf{w}^H\mathbf{B}_3\mathbf{w}.\nonumber
\end{align}

The above problem is SOCP and can be solved by CVX.
Furthermore,
it is obvious
that the 
 problem (P11) can also be efficiently solved by using
 Lemma \ref{lem:complex_Convex_trust_region problem}.

Next,
we introduce the new definitions
\begin{small}
\begin{align}
&
c_6
\triangleq 
(\mathbf{w}^H\mathbf{B}_5\mathbf{f}\mathbf{f}^H\mathbf{B}_5^H\mathbf{w})
/(2\rho)
\!+\!
\Re\{\lambda_2^{\ast}\mathbf{w}^H\mathbf{B}_5\mathbf{f} \}-\mathbf{w}^H\mathbf{B}_3\mathbf{w},\\
&a_1
\triangleq 
1/(2\rho),
a_2
\triangleq 
-\mathbf{w}^H\mathbf{B}_5\mathbf{f}/\rho - \Re\{\lambda_2^{\ast} \},
a_4
\triangleq 
\mathbf{w}^H\mathbf{B}_5\mathbf{f}\mathbf{f}^H\mathbf{B}_5^H\mathbf{w}.\nonumber
\end{align}
\end{small}

The problem 
 w.r.t. $b$ is written as
\begin{subequations}
\begin{align}
\textrm{(P12)}:
\mathop{\textrm{min}}
\limits_{
b
}\
&
a_1b^2 + a_2b+c_6+\frac{a_4}{b}
 \label{P12_obj}
\end{align}
\end{subequations}

The problem (P12) is an unconstrained convex problem 
and its optimal solution can be obtained 
by setting its derivation to zero.

The inner layer of the PDD method updates 
$\mathbf{w}$, 
$\mathbf{f}$ 
and $b$ in a BCD manner until convergence.

\emph{\textit{Outer Layer Procedure}}

When the inner layer achieves convergence,
the outer layer will selectively update
the value of
dual variables 
$\{\bm{\lambda}_1, \lambda_2 \}$
or the penalty coefficient $\rho$.
Specifically,
\begin{enumerate}[1)]
\item
when the equations 
$\mathbf{w} = \mathbf{f}$ 
and 
$\mathbf{w}^H \mathbf{B}_{5} \mathbf{f} = b$
are nearly satisfied,
the
Lagrangian multiplies $\{\bm{\lambda}_1, \lambda_2 \}$  will
 be  respectively updated in a gradient ascent manner as follows:
 \begin{subequations}
 \begin{align}
 &\bm{\lambda}_{1}^{(k+1)}:=\bm{\lambda}_{1}^{(k)}
 +\rho^{-1}(\mathbf{w} -\mathbf{f}),\\
 & \lambda_2^{(k+1)}:= \lambda_2^{(k)}
 +\rho^{-1}(\mathbf{w}^H \mathbf{B}_{5} \mathbf{f} - b).
 \end{align}
 \end{subequations}
 
 \item
 when the equality constraints $\mathbf{w} = \mathbf{f}$ 
and/or
$\mathbf{w}^H \mathbf{B}_{5} \mathbf{f} = b$
are not achieved,
to force
 $\mathbf{w} = \mathbf{f}$ 
and/or
$\mathbf{w}^H \mathbf{B}_{5} \mathbf{f} = b$
to
be approached in the subsequent iterations,
the outer layer will update 
 the penalty parameter $\rho^{-1}$
 as follows:
 \begin{align}
 \big(\rho^{(k+1)}\big)^{-1}:=c^{-1}\cdot\big(\rho^{(k)}\big)^{-1},
\end{align}
where
the positive constant $c$ is usually smaller than 1 and 
is often selected within the range of [0.8, 0.9].
\end{enumerate}

The previously developed algorithm based on the PDD framework to solve the problem (P3) is summarized in Alg.\ref{alg:PDD}.
\begin{algorithm}[t]
\caption{PDD Method to Solve (P3)}
\label{alg:PDD}
\begin{algorithmic}[1]
\STATE {initialize}
$\mathbf{w}^{(0)}$,
$\mathbf{f}^{(0)}$,
$b^{(0)}$,
$\bm{\lambda}_{1}^{(0)}$,
${\lambda}_{2}^{(0)}$,
${\rho}^{(0)}$
and
$k=1$;
\REPEAT
\STATE set $\mathbf{w}^{(k-1,0)}:=\mathbf{w}^{(k-1)}$,
$\mathbf{f}^{(k-1,0)}:=\mathbf{f}^{(k-1)}$,
$b^{(k-1,0)}:=b^{(k-1)}$,
$t=0$;
\REPEAT
\STATE  update $\mathbf{w}^{(k-1,t+1)}$  by  solving (P7);
\STATE  update $\mathbf{f}^{(k-1,t+1)}$  by  solving (P11);
\STATE  update $b^{(k-1,t+1)}$  by  solving (P12);
\STATE  $t++$;
\UNTIL{$convergence$}
\STATE set $\mathbf{w}^{(k)}:=\mathbf{w}^{(k-1,\infty)}$,
$\mathbf{f}^{(k)}:=\mathbf{f}^{(k-1,\infty)}$,
$b^{(k)}:=b^{(k-1,\infty)}$;
\IF{$\Vert\mathbf{w}^{(k)}-\mathbf{f}^{(k)}\Vert_{\infty}\leq\eta_k $
and
$\vert(\mathbf{w}^{(k)})^H \mathbf{B}_{5} \mathbf{f}^{(k)} - b^{(k)}\vert\leq\eta_k $   }
\STATE{ $\bm{\lambda}_{1}^{(k+1)}:=\bm{\lambda}_{1}^{(k)}
+\dfrac{1}{{\rho}^{(k)}}(\mathbf{w}^{(k)}-\mathbf{f}^{(k)})$, 
${\lambda}_{2}^{(k+1)}:={\lambda}_{2}^{(k)}
+\dfrac{1}{{\rho}^{(k)}}((\mathbf{w}^{(k)})^H \mathbf{B}_{5} \mathbf{f}^{(k)} - b^{(k)})$, 
${\rho}^{(k+1)}:= {\rho}^{(k)}$};
\ELSE
\STATE{$\bm{\lambda}_{1}^{(k+1)}:=\bm{\lambda}_{1}^{(k)}$,
${\lambda}_{2}^{(k+1)}:={\lambda}_{2}^{(k)}$,
 $1/{\rho}^{(k+1)}:= 1/(c\cdot{\rho}^{(k)})$};
\ENDIF \STATE $k++$;
\UNTIL{$\Vert\mathbf{w}^{(k)}-\mathbf{f}^{(k)}\Vert_2$
and
$\vert(\mathbf{w}^{(k)})^H \mathbf{B}_{5} \mathbf{f}^{(k)} - b^{(k)}\vert$
are sufficiently small simultaneously;}
\end{algorithmic}
\end{algorithm}

\subsection{Updating The Receiver Filter $\{\mathbf{u}_k\}$}

In this subsection, 
we investigate the optimization of 
the user receive filter $\{\mathbf{u}_k\}$ 
while keeping other variables fixed.
By introducing the new coefficients as follows

\begin{small}
\begin{align}
 & \mathbf{d}_{1,k}
  \!\triangleq\!\!
  \sqrt{(1\!+\!\gamma_k)}\omega_k^{\ast}\sqrt{{q}_k}\mathbf{h}_k,
  c_7
  \!\!\triangleq\!\!
  {\sum}_{k=1}^{K}\!( \textrm{log}(1\!+\!\gamma_k)\! -\! \gamma_k  )
  \!\!-\!\!{R}_{t},\\
  & \mathbf{D}_{1,k}
 \!\!\triangleq\!\!
 \vert\omega_k\vert^2
 (\!
 {\sum}_{i =1}^{K} {q}_i\vert\mathbf{u}_k^H\mathbf{h}_i\vert^2 
 \!\!+\!\! \vert\alpha\vert^2\!\mathbf{a}_{r}\mathbf{a}_{t}^H\!\mathbf{W}
 \mathbf{W}^H\!\mathbf{a}_{t}\mathbf{a}_{r}^H
 \!\!+\!\!\sigma^2\Vert\mathbf{u}_k^H\Vert^2_2
 ),\nonumber
\end{align}
\end{small}

\noindent
the constraint (\ref{P2_c1})
is rewritten as
\begin{align}
{\sum}_{k=1}^{K}
(
\mathbf{u}_k^H \mathbf{D}_{1,k} \mathbf{u}_k
- 2\Re\{ \mathbf{u}_k^H\mathbf{d}_{1,k} \}
)
- c_7
\leq 
0
\end{align}

Therefore,
the optimization of  $\{\mathbf{u}_k\}$
is reduced to solve the following feasibility characterization problem as
\begin{subequations}
\begin{align}
\textrm{(P13)}:
&\mathop{\textrm{Find}}
\limits_{
\{\mathbf{u}_k\}
}\
\{\mathbf{u}_k\}
 \label{P13_obj}\\
\textrm{s.t.}\ &
{\sum}_{k=1}^{K}
\!(\!
\mathbf{u}_k^H \mathbf{D}_{1,k} \mathbf{u}_k
\!\!-\!\! 2\Re\{ \mathbf{u}_k^H\mathbf{d}_{1,k} \}
\!)
\!\!-\!\! c_7
\!\leq \!
0
.\label{P13_c1}
\end{align}
\end{subequations}

The problem (P13), also referred to as the Phase-I problem \cite{ref_Convex Optimization}, 
is addressed by solving another closely related problem as follows:
\begin{subequations}
\begin{align}
\textrm{(P14)}:
&\mathop{\textrm{min}}
\limits_{
\{\mathbf{u}_k\},
\alpha_1
}\
\alpha_1
 \label{P14_obj}\\
&\textrm{s.t.}\ 
{\sum}_{k=1}^{K}
\!(\!
\mathbf{u}_k^H \mathbf{D}_{1,k} \mathbf{u}_k
\!\!-\!\! 2\Re\{ \mathbf{u}_k^H\mathbf{d}_{1,k} \}
\!)
\!\!-\!\! c_7
\!\leq \!
\alpha_1
.\label{P14_c1}
\end{align}
\end{subequations}

As stated in
\cite{ref_Convex Optimization}, \cite{ref_Phase},
minimizing (P14) is to identify more  
``feasible" 
$\{\mathbf{u}_k\}$,
which offer a greater margin for satisfying  the constraint (\ref{P13_c1}) 
and thereby facilitate the optimization of other variables.
Clearly,
optimality in problem (P14)
is achieved only when equality 
is reached in constraint (\ref{P14_c1}).
Consequently, 
solving (P14) equates 
to minimize the expression on the left side of (\ref{P14_c1}), 
which is given as
\begin{subequations}
\begin{align}
\textrm{(P15)}:
&\mathop{\textrm{min}}
\limits_{
\{\mathbf{u}_k\}
}\
{\sum}_{k=1}^{K}
\!(
\mathbf{u}_k^H \mathbf{D}_{1,k} \mathbf{u}_k
\!\!-\!\! 2\Re\{ \mathbf{u}_k^H\mathbf{d}_{1,k} \}
)
\!\!-\!\! c_7
 \label{P15_obj}
\end{align}
\end{subequations}

Furthermore,
the problem (P15)
can be  decomposed into $K$ independent sub-problems
with each sub-problem defined as follows:
\begin{subequations}
\begin{align}
\textrm{(P15$_k$)}:
&\mathop{\textrm{min}}
\limits_{
\mathbf{u}_k
}\
\mathbf{u}_k^H \mathbf{D}_{1,k} \mathbf{u}_k
- 2\Re\{ \mathbf{u}_k^H\mathbf{d}_{1,k} \}
 \label{P15_obj_k}
\end{align}
\end{subequations}

Since (P15$_k$) is an  
unconstrained convex quadratic problem,
the optimal solution $\mathbf{u}_k^{\star}$ 
can be determined by setting its derivative to zero,
which is formulated as
\begin{align}
\mathbf{u}_k^{\star} = \mathbf{D}_{1,k}^{-1}\mathbf{d}_{1,k},
\forall k \in \mathcal{K}.\label{UU_closed_solution}
\end{align}

\subsection{Optimizing The User Transmission Power}

After fixing other variables,
the problem w.r.t. $\{q_k\}$
can be expressed as follows
\begin{subequations}
\begin{align}
\textrm{(P16)}:
&\mathop{\textrm{Find}}
\limits_{
\{q_k\}
}\
\{q_k\}
 \label{P16_obj}\\
 &\textrm{s.t.}\
 {\sum}_{k=1}^{K}
 ( a_{5,k}q_k - a_{6,k}\sqrt{q_k}  )-c_8\leq  0, \\
 & 0 \leq q_k \leq P_{u,k}, \forall k \in \mathcal{K},
\end{align}
\end{subequations}
with the newly introduced coefficients specified as follows
\begin{small}
\begin{align}
&a_{5,i}
\!\triangleq\!\!
{\sum}_{k=1}^{K}\vert\omega_k\vert^2
\vert\mathbf{u}_k^H\mathbf{h}_i\vert^2 ,
a_{6,k}
\!\triangleq\!\!
2\sqrt{(1+\gamma_k)}
\Re\{\omega_k^{\ast}\mathbf{u}_k^H\mathbf{h}_k \},\\
& c_8\!
\!\triangleq\!\!
{\sum}_{k=1}^{K}\!\!
\big(\!
\textrm{log}(1\!\!+\!\!\gamma_k)
\!-\!
\gamma_k
\!-\!\!
\vert\omega_k\vert^2\!
(\!\Vert\! \mathbf{u}_k^H \! \alpha
\!\mathbf{a}_{r}\mathbf{a}_{t}^H\!\mathbf{W}\!\Vert^2_2
\! +\! \sigma^2\!\Vert\mathbf{u}_k^H\!\Vert^2_2\!
)\!
\big)
\!\!-\!\!
R_t.\nonumber
\end{align}
\end{small}

Note that the problem (P16)
is also a feasibility characterization problem.
By introducing an auxiliary variable $\alpha_2$,
(P16) can be rewritten as
\begin{subequations}
\begin{align}
\textrm{(P17)}:
&\mathop{\textrm{min}}
\limits_{
\{q_k\},
\alpha_2
}\
\alpha_2
 \label{P17_obj}\\
 &\textrm{s.t.}\
 {\sum}_{k=1}^{K}
 ( a_{5,k}q_k - a_{6,k}\sqrt{q_k}  )-c_8\leq  \alpha_2, \\
 & 0 \leq q_k \leq P_{u,k}, \forall k \in \mathcal{K},
\end{align}
\end{subequations}

Furthermore,
following the arguments 
(\ref{P13_obj})-(\ref{P15_obj}) 
of the above subsection,
we turn to solve the following problem
\begin{subequations}
\begin{align}
\textrm{(P18)}:
&\mathop{\textrm{min}}
\limits_{
\{q_k\}
}\
{\sum}_{k=1}^{K}
 ( a_{5,k}q_k - a_{6,k}\sqrt{q_k}  )
 \label{P18_obj}\\
  &\textrm{s.t.}\
  0 \leq q_k \leq P_{u,k}, \forall k \in \mathcal{K},
\end{align}
\end{subequations}
which problem
(P18) is convex and can be solved by CVX.
Next, 
we will develop the closed-form solution to
update $\{q_k\}$.
Obviously,
(P18)
can also be  decomposed into $K$ independent sub-problems,
which the sub-problem is formulated as 
\begin{subequations}
\begin{align}
\textrm{(P18$_k$)}:
&\mathop{\textrm{min}}
\limits_{
q_k
}\
  a_{5,k}q_k - a_{6,k}\sqrt{q_k}  
 \label{P18_obj_k}\\
   &\textrm{s.t.}\
  0 \leq q_k \leq P_{u,k},
\end{align}
\end{subequations}

Furthermore,
we define 
$p_k \triangleq \sqrt{q_k}$
and 
$\bar{P}_k \triangleq \sqrt{P_{u,k}} $,
and then the problem 
(P18$_k$) is rewritten as
\begin{subequations}
\begin{align}
\textrm{(P19)}:
&\mathop{\textrm{min}}
\limits_{
p_k
}\
  a_{5,k}p_k^2 - a_{6,k}{p_k}  
 \label{P19_obj}\\
   &\textrm{s.t.}\
  0 \leq p_k \leq \bar{P}_k.
\end{align}
\end{subequations}

Since 
$a_{5,k} > 0$,
the closed solution can be determined by 
judge the position of the axis of symmetry,
which is given as
\begin{align}
p_k^{\star}=
\left\{
\begin{aligned}
&0,         \    p_{k,x} < 0,\\
&\bar{P}_k, \   p_{k,x} > \bar{P}_k,\\
&p_{k,x},   \  \textrm{otherwise},
\end{aligned}
\right.\label{p_star}
\end{align}
where 
$p_{k,x} \triangleq \frac{a_{6,k}}{2a_{5,k}}  $
and $q_k^{\star} = (p_k^{\star})^2 $.

\subsection{Optimizing The Position of BS FA}

With fixed other variables,
the update for the position vector of the BS FA, 
i.e., $\mathbf{d}_r$,
can be formulated as follows
\begin{subequations}
\begin{align}
\textrm{(P20)}:
\mathop{\textrm{min}}
\limits_{
\mathbf{d}_r
}& \
\mathbf{d}_r^T\mathbf{A}_{4,3}\mathbf{d}_r
+
\mathbf{d}_r^T\mathbf{b}_4 - c_{16}
 \label{P20_obj}\\
\textrm{s.t.}\ 
& {\sum}_{k=1}^{K}
(
\mathbf{\bar{h}}_{1,k}^H
\mathbf{A}_{6,k}
\mathbf{\bar{h}}_{1,k}
+
\Re\{
\mathbf{b}_{7,k}^H\mathbf{\bar{h}}_{1,k}
\}
)\\
&+
\mathbf{a}_r^H\mathbf{A}_9\mathbf{a}_r
-c_{20}
\leq 0,\nonumber\\
& d_{r,1} \geq 0,
d_{r,N_r} \leq d_{max},\\
&d_{r,n}\!-\!d_{r,n-1}\! \geq\! d_{min},
n\! =\! 2,3, \cdots, N_r.\label{P20_c3}
\end{align}
\end{subequations}
where 
the above newly introduced notions are given in (\ref{MA_BS_1}).
\begin{figure*}
\begin{small}
\begin{align}
&
\mathbf{\bar{h}}_{1,k}
\triangleq 
\textrm{vec}
(
\mathbf{H}_{0,k}(\mathbf{d}_{r})
),
\mathbf{D}_r
\triangleq 
j\frac{2\pi}{\lambda}\textrm{diag}(\mathbf{d}_r)\cos(\theta_0),
\mathbf{D}_t
\triangleq 
j\frac{2\pi}{\lambda}\textrm{diag}(\mathbf{d}_t)\cos(\theta_0),
\mathbf{A}_t
\triangleq 
\mathbf{a}_{t}\mathbf{a}_{t}^H,
\mathbf{W}_t
\triangleq 
\mathbf{W}\mathbf{W}^H,\label{MA_BS_1}\\
&
\mathbf{A}_{4,1}
\triangleq 
(\frac{2\pi}{\lambda})^2\mathbf{I}
\textrm{tr}( 
\mathbf{A}_t\mathbf{W}_t ),
\mathbf{b}_1
\triangleq 
\textrm{tr}(
\mathbf{A}_t
\mathbf{D}_t^H
\mathbf{W}_t)
(-j\frac{2\pi}{\lambda}\cos{\theta_0}
\mathbf{1}),
\mathbf{b}_2
\triangleq 
\textrm{tr}
(\mathbf{D}_t
\mathbf{A}_t
\mathbf{W}_t)
(j\frac{2\pi}{\lambda}\cos{\theta_0}
\mathbf{1}),
c_{13}
\triangleq 
\textrm{tr}
(N_r
\mathbf{D}_t
\mathbf{A}_t
\mathbf{D}_t^H
\mathbf{W}_t),\nonumber\\
&
\mathbf{b}_3
\triangleq 
(
-j\frac{2\pi}{\lambda}\cos{\theta_0}
\mathbf{1}
)
\textrm{tr}
(
\mathbf{A}_t
\mathbf{W}_t
),
c_{14}
\triangleq 
\textrm{tr}
(
N_r\mathbf{D}_t
\mathbf{A}_t
\mathbf{W}_t
),
c_{15}
\triangleq 
\textrm{tr}(\mathbf{A}_3\mathbf{W}_t),
\mathbf{A}_{4,2}
\!\triangleq\!
 \mathbf{b}_3\mathbf{b}_3^T/c_{15},
\mathbf{b}_4
\!\triangleq \!
2c_14\mathbf{b}_3/c_{15}\!-\!(\mathbf{b}_1\! +\! \mathbf{b}_2),\nonumber\\
&c_{16}
\!\triangleq \!
-\vert c_{14}\vert^2/c_{15}\! +\! c_{13},
\mathbf{A}_{4,3} 
\!\triangleq \!
\mathbf{A}_{4,2}\! -\! \mathbf{A}_{4,1},
\mathbf{A}_{9}
\!\triangleq \!
{\sum}_{k=1}^{K}\vert\omega_k\vert^2
\big(
(\mathbf{a}_t^H\mathbf{W}_t\mathbf{a}_t)^T
\!\!\otimes\!
\mathbf{u}_k\mathbf{u}_k^H\big),
\mathbf{A}_{6,i}
\!\triangleq \!\!\!
{\sum}_{k=1}^{K}\!
\vert \omega_k \vert^2\!
(q_i \mathbf{u}_k\mathbf{u}_k^H )^T\!
\!\!\otimes \!
(\!
\bm{\Sigma}_i^H\! \mathbf{h}_{0,i}\mathbf{h}_{0,i}^H \bm{\Sigma}_i
\!
),\nonumber\\
&\mathbf{b}_{7,k}
\!\!\triangleq \!\!
-\!2 \sqrt{\!1\!+\!\gamma_k}
(\mathbf{u}_k^T\! 
\!\otimes\!
\omega_k \sqrt{q_k}\mathbf{h}_{0,k}^H\bm{\Sigma_k}),
c_{20}
\!\triangleq \!\!\!
{\sum}_{k=1}^{K}\!
\big(
\textrm{log}(1\!\!+\!\!\gamma_k)
\!\!-\!\!
\gamma_k
\!\!-\!\!
\vert \omega_k \vert^2
\!(
\sigma^2\Vert\mathbf{u}_k^H\Vert_2^2
)
\big)
\!\!-\!\!R_t.\nonumber
\end{align}
\end{small}
\bm{\hrule}
\end{figure*}

Note that the problem (P20) is non-convex.
We proceed to update the positions of the FAs one by one.
The sub-problem for optimizing $n$-th FA's position $d_{r,n}$
is given as
\begin{subequations}
\begin{align}
\textrm{(P21)}:
&\mathop{\textrm{min}}
\limits_{
{d}_{r,n}
} \
a_{7,n}d_{r,n}^2
+
b_{6,n}d_{r,n}
-c_{19,n}
 \label{P21_obj}\\
\textrm{s.t.}\ 
& {\sum}_{k=1}^{K}\!
(
\mathbf{h}_{1,k,n}^H
\mathbf{A}_{7,k,n,n}
\mathbf{h}_{1,k,n}
\!\!+\!\!
\Re\{
\mathbf{b}_{10,k,n}^H\mathbf{h}_{1,k,n}\!\} \label{P21_c1} \\
&+c_{21,k,n}
+c_{22,k,n}
)
+\Re
\{
b_{15,n}^{\ast}\mathbf{a}_r[n]
\}+c_{27,n}\leq 0,\nonumber\\
&   d_{r,n}-d_{r,n-1} \geq  d_{min},  \label{P21_c2}\\
& d_{r,n+1} -  d_{r,n}\geq  d_{min}, \label{P21_c3}\\
& 0 \leq  d_{r,n} \leq d_{max},\label{P21_c4}
\end{align}
\end{subequations}
where 
the new coefficients are defined in (\ref{MA_BS_2}).
\begin{figure*}
\begin{align}
&a_{7,n}
\triangleq
\mathbf{A}_3[n,n],
b_{6,n}
\triangleq
2{\sum}_{i\neq n}^{N_r}
d_{r,i}\mathbf{A}_3[i,n]+\mathbf{b}_4[n],
c_{19,n}
\triangleq
{\sum}_{i\neq n}^{N_r}{\sum}_{j\neq n}^{N_r}
d_{r,i}\mathbf{A}_3[i,j]d_{r,j}
-
{\sum}_{i\neq n}^{N_r}
d_{r,i}\mathbf{b}_4[i]
+c_{16}, \label{MA_BS_2} \\
&
\mathbf{A}_{7,k,n,m}
\triangleq
\mathbf{A}_{6,k}
[
(n-1)L_{r,k}+1:nL_{r,k},
(m-1)L_{r,k}+1:mL_{r,k}
],
\mathbf{b}_{8,k,n}
\triangleq
\mathbf{b}_{7,k}
[
(n-1)L_{r,k}+1:nL_{r,k}
],\nonumber\\
&\mathbf{b}_{9,k,n}
\!\triangleq\!
2\!{\sum}_{m\neq n}^{N_r}
\!\mathbf{A}_{7,k,m,n}^H\mathbf{h}_{1,k,m},
c_{21,k,n}
\!\triangleq\!
{\sum}_{i\neq n}^{N_r}
{\sum}_{j\neq n}^{N_r}
\mathbf{h}_{1,k,i}^H\mathbf{A}_{7,k,i,j}\mathbf{h}_{1,k,j},
c_{22,k,n}
\!\triangleq\!
{\sum}_{i\neq n}^{N_r}
\Re\{
\mathbf{b}_{8,k,i}^H\mathbf{h}_{1,k,i}
\},\nonumber\\
&\mathbf{b}_{10,k,n}
\!\triangleq\!
\mathbf{b}_{8,k,n}
\!+\!
\mathbf{b}_{9,k,n},
b_{15,n}
\triangleq
2{\sum}_{i \neq n}^{N_r}
a_r[i]\mathbf{A}_{9,k}[i,n]^{\ast},
c_{27,n}
\triangleq
\mathbf{A}_{9,k}[n,n]
{\sum}_{i \neq n}^{N_r}
{\sum}_{j \neq n}^{N_r}
a_r[i]^{\ast}
\mathbf{A}_{9,k}[i,j]
a_r[j].\nonumber
\end{align}
\bm{\hrule}
\end{figure*}
Next,
for simplicity,
we define $d_{r,0}
\triangleq
-d_{min}$,
$d_{r,N_r+1}
\triangleq
d_{max}+d_{min}$,
and the  constraints (\ref{P21_c2})$-$(\ref{P21_c4}) can be rewritten as 
\begin{align}
&\left\{
\begin{aligned}
&d_{r,n}-d_{r,n-1} \geq  d_{min}\\
&d_{r,n+1} -  d_{r,n}\geq  d_{min}\\
& 0 \leq  d_{r,n} \leq d_{max}\\
\end{aligned}
\right.\label{P21_c5}\\
&\Longleftrightarrow
d_{r,n-1} +d_{min} \leq d_{r,n} \leq d_{r,n+1}-d_{min}.\nonumber
\end{align}

Therefore, 
the problem (P21) is rewritten by
\begin{subequations}
\begin{align}
\textrm{(P22)}:
&\mathop{\textrm{min}}
\limits_{
{d}_{r,n}
} \
a_{7,n}d_{r,n}^2
+
b_{6,n}d_{r,n}
-c_{19,n}
 \label{P22_obj}\\
\textrm{s.t.}\ 
& {\sum}_{k=1}^{K}\!
(
\mathbf{h}_{1,k,n}^H
\mathbf{A}_{7,k,n,n}
\mathbf{h}_{1,k,n}
\!\!+\!\!
\Re\{
\mathbf{b}_{10,k,n}^H\mathbf{h}_{1,k,n}\!\} \label{P22_c1} \\
&+c_{21,k,n}
+c_{22,k,n}
)
+\Re
\{
b_{15,n}^{\ast}\mathbf{a}_r[n]
\}+c_{27,n}\leq 0,\nonumber\\
&d_{r,n-1} +d_{min} \leq d_{r,n} \leq d_{r,n+1}-d_{min},
\end{align}
\end{subequations}

Since the constraint (\ref{P22_c1})  is non-convex,
we adopt the MM method to conduct a convex upper bound.
Firstly,
$d_{r,n,0}$ defines the value obtained from the previous iteration,
$\lambda_{max}(\mathbf{A}_{7,k,n,n})$ is the largest eigenvalue of 
matrix $\mathbf{A}_{7,k,n,n}$
and
$\mathbf{h}_{1,k,n,0}
\triangleq
\mathbf{h}_{1,k,n}(d_{r,n,0})
$.
The upper bound of the equation
$\mathbf{h}_{1,k,n}^H
\mathbf{A}_{7,k,n,n}
\mathbf{h}_{1,k,n}
+
\Re\{
\mathbf{b}_{10,k,n}^H\mathbf{h}_{1,k,n}\}$
is presented in (\ref{MA_BS_MM_1}),
\begin{figure*}
\begin{small}
\begin{align}
&\mathbf{h}_{1,k,n}^H
\mathbf{A}_{7,k,n,n}
\mathbf{h}_{1,k,n}
+
\Re\{
\mathbf{b}_{10,k,n}^H\mathbf{h}_{1,k,n}\} \label{MA_BS_MM_1} \\
&=
(\mathbf{h}_{1,k,n}\!\!-\!\mathbf{h}_{1,k,n,0})^H
\mathbf{A}_{7,k,n,n}
(\mathbf{h}_{1,k,n}\!\!-\!\mathbf{h}_{1,k,n,0})
\!+\!
2\Re\{
\mathbf{h}_{1,k,n,0}^H\mathbf{A}_{7,k,n,n}
(\mathbf{h}_{1,k,n}\!-\!\mathbf{h}_{1,k,n,0})
\}
\!\!+\!\!
\mathbf{h}_{1,k,n,0}^H
\mathbf{A}_{7,k,n,n}
\mathbf{h}_{1,k,n,0}
\!\!+\!\!
\Re\{
\mathbf{b}_{10,k,n}^H\mathbf{h}_{1,k,n}\}\nonumber\\
&
\leq
\lambda_{max}(\mathbf{A}_{7,k,n,n})
\Vert\mathbf{h}_{1,k,n}-\mathbf{h}_{1,k,n,0}\Vert_2^2
+
2\Re\{
\mathbf{h}_{1,k,n,0}^H\mathbf{A}_{7,k,n,n}
(\mathbf{h}_{1,k,n}-\mathbf{h}_{1,k,n,0})
\}
+
\mathbf{h}_{1,k,n,0}^H
\mathbf{A}_{7,k,n,n}
\mathbf{h}_{1,k,n,0}
+
\Re\{
\mathbf{b}_{10,k,n}^H\mathbf{h}_{1,k,n}\}\nonumber\\
&
=\Re\{
\mathbf{b}_{12,k,n}^H\mathbf{h}_{1,k,n}
\}+c_{23,k,n},\nonumber
\end{align}
\end{small}
\bm{\hrule}
\end{figure*}
where
$ \mathbf{b}_{11,k,n} 
\triangleq
 -2 \lambda_{max}(\mathbf{A}_{7,k,n,n})
\mathbf{h}_{1,k,n,0}
 +\mathbf{A}_{7,k,n,n}^H\mathbf{h}_{1,k,n,0}  $,
 $
 c_{23,k,n}
 \triangleq
 2\lambda_{max}(\mathbf{A}_{7,k,n,n})L_{r,k}
 -
 ( \mathbf{h}_{1,k,n,0}^H\mathbf{A}_{7,k,n,n}\mathbf{h}_{1,k,n,0}  )^{\ast}
 $
 and
 $ \mathbf{b}_{12,k,n} 
 \triangleq
 \mathbf{b}_{10,k,n} 
 +
 \mathbf{b}_{11,k,n} 
 $,
 and then,
 by using 
 the second-order Taylor expansion again,
 we can  construct the convex surrogate
functions of 
$\Re\{
\mathbf{b}_{12,k,n}^H\mathbf{h}_{1,k,n}
\}$
and
$\Re
\{
b_{15,n}\mathbf{a}_r[n]
\}$ 
in
(\ref{MA_BS_MM_2})
and
(\ref{MA_BS_MM_3}),
respectively,
where
\begin{small}
\begin{align}
&
\nabla h_{3,k,n,0}
\triangleq
\frac{\partial h_{3,k,n,0}}{\partial d_{r,n,0}}
=
-
{\sum}_{i=1}^{L_{r,k}}
\vert\mathbf{b}_{12,k,n}^{\ast}[i]\vert
\frac{2\pi}{\lambda}\sin(\theta_{k,i}^r)\\
&\sin( \frac{2\pi}{\lambda}\sin(\theta_{k,i}^r) 
- \angle\mathbf{b}_{12,k,n}^{\ast}[i]  ),
 \tau_{2,k,n}
 \triangleq
 \frac{4\pi^2}{\lambda^2}
 {\sum}_{i=1}^{L_{r,k}}
 \vert \mathbf{b}_{12,k,n}^{\ast}[i] \vert,\nonumber\\
  &\nabla \tilde{a}_{r}(d_{r,n,0})
 \triangleq
 -\vert b_{15,n}^{\ast} \vert
 \frac{2\pi}{\lambda}\sin(\theta_0)
 \sin( \frac{2\pi}{\lambda}\sin(\theta_0) 
 - \angle b_{15,n}^{\ast} ),\nonumber\\
 &\tau_{4,n}
 \triangleq
 \frac{4\pi^2}{\lambda^2}\vert b_{15,n}^{\ast} \vert,
b_{13,k,n}
\triangleq
\nabla h_{3,k,n,0}- \tau_{2,k,n} d_{r,n,0},\nonumber\\
&
c_{25,k,n}
\triangleq
\frac{1}{2}\tau_{2,k,n}d_{r,n,0}^2
-\nabla h_{3,k,n,0}d_{r,n,0}
+
 h_{3,k,n,0},\nonumber\\
 &
 b_{16,n} 
 \triangleq
 \nabla \tilde{a}_{r}(d_{r,n,0})
 -
 \tau_{4,n}d_{r,n,0}.\nonumber
\end{align}
\end{small}

\begin{figure*}
\begin{small}
\begin{align}
&\Re\{
\mathbf{b}_{12,k,n}^H\mathbf{h}_{1,k,n}
\}
=
{\sum}_{i=1}^{L_{r,k}}
\vert {b}_{12,k,n}^{\ast}[i] \vert
\cos
(
\frac{2\pi}{\lambda}d_{r,n}\sin(\theta_0)
-\angle{b}_{12,k,n}^{\ast}[i]
)
\triangleq
h_{3,k,n} \label{MA_BS_MM_2} \\
&
\leq
h_{3,k,n,0}(d_{r,n,0})
+
\nabla h_{3,k,n,0}^T(d_{r,n}-d_{r,n,0})
+
\frac{1}{2}\tau_{2,k,n}
(d_{r,n}-d_{r,n,0})(d_{r,n}-d_{r,n,0})
=\frac{1}{2}\tau_{2,k,n}d_{r,n}^2
+b_{13,k,n}d_{r,n} + c_{25,k,n},\nonumber\\
&
\Re
\{
b_{15,n}^{\ast}\mathbf{a}_r[n]
\}
=
\vert b_{15,n}^{\ast}\vert
\cos(
\frac{2\pi}{\lambda}d_{r,n}\sin(\theta_0)
-
\angle b_{15,n}^{\ast}
)
\triangleq
\tilde{a}_{r}(d_{r,n})\label{MA_BS_MM_3}\\
&\leq
\tilde{a}_{r}(d_{r,n,0})
+
\nabla \tilde{a}_{r}(d_{r,n,0})(d_{r,n}-d_{r,n,0})
+
\frac{1}{2}\tau_{4,n}(d_{r,n}-d_{r,n,0})(d_{r,n}-d_{r,n,0})
=\frac{1}{2}\tau_{4,n}d_{r,n}^2
+b_{16,n}d_{r,n}+c_{28,n}\nonumber
\end{align}
\end{small}
\bm{\hrule}
\end{figure*}

Based on the previous transformation,
the problem w.r.t. $d_{r,n}$
can be written as
\begin{subequations}
\begin{align}
\textrm{(P23)}:
&\mathop{\textrm{min}}
\limits_{
{d}_{r,n}
} \
a_{7,n}d_{r,n}^2
+
b_{6,n}d_{r,n}
-c_{19,n}
 \label{P23_obj}\\
\textrm{s.t.}\ 
& 
\tau_3d_{r,n}^2
+b_{14,n}d_{r,n} + c_{26,n}
\leq 0,\label{P23_c1}\\
&
d_{r,n-1} +d_{min} \leq d_{r,n} \leq d_{r,n+1}-d_{min},\label{P23_c2}
\end{align}
\end{subequations}
where 
$\tau_3 \triangleq
{\sum}_{k=1}^{K}
\frac{1}{2}\tau_{2,k,n}
+
\frac{1}{2}\tau_{4,n}
 $,
 $
 b_{14,n}
 \triangleq
 {\sum}_{k=1}^{K}b_{13,k,n}
 +
 b_{16,n}
 $
 and
 $c_{26,n}$ is a constant.
 The problem (P23) is convex and can be solved by CVX.
 Furthermore,
we proceed to present its analytic solution.
Firstly,
according to the structure of the constraints (\ref{P23_c1})$-$(\ref{P23_c2}),
the feasible area of (P23) can be rewritten as:
\begin{align}
d_{r,n}\in
[
d_{r,n,low}&\triangleq
\textrm{max}\{ d_{r,n,r1}, d_{r,n-1} +d_{min} \},\label{MA_BS_dd}\\
d_{r,n,up}&\triangleq
\textrm{min}\{ d_{r,n,r2}, d_{r,n+1} -d_{min} \}
],\nonumber
\end{align}
where
$d_{r,n,r1}$
and
$d_{r,n,r2}$
denote the roots of the equation
$\tau_3d_{r,n}^2
+b_{14,n}d_{r,n} + c_{26,n}
= 0$ with $d_{r,n,r1} \leq d_{r,n,r2}$.
The proof details of (\ref{MA_BS_dd}) can be seen in the Appendix B.
Therefore,
(P23) is rewritten as
\begin{subequations}
\begin{align}
\textrm{(P24)}:
&\mathop{\textrm{min}}
\limits_{
{d}_{r,n}
} \
a_{7,n}d_{r,n}^2
+
b_{6,n}d_{r,n}
-c_{19,n}
 \label{P24_obj}\\
\textrm{s.t.}\ 
& 
d_{r,n}\in
[
d_{r,n,low},
d_{r,n,up}
].
\end{align}
\end{subequations}

And then,
the closed solution of (P24) is presented 
in the following theorem.
\begin{theorem}\label{alg:MA_BS_closed_solution}
We define $d_{r,n,sym}$  as the axis of symmetry of 
the  function $ f(d_{r,n}) \triangleq  a_{7,n}d_{r,n}^2
+
b_{6,n}d_{r,n}
-c_{19,n}$.
The
optimal values of $d_{r,n}$ are obtained according to one of the
following three cases:
\begin{itemize} \label{Throrem_1}
\item[] \underline{CASE-I}:
if $a_{7,n} = 0$,
by investigating the value of $b_{6,n}$,
the optimal value of $d_{r,n}$ is given as
\begin{align}
d_{r,n}^{\star}=
\left\{
\begin{aligned}
&d_{r,n,low},\ b_{6,n} >0\\
&d_{r,n,up},\ b_{6,n} <0\\
& \text{arbitrary value in}\  [
d_{r,n,low},
d_{r,n,up}
],\ b_{6,n} =0
\end{aligned}
\right.
\end{align}

\item[] \underline{CASE-II}:
when
$a_{7,n} > 0$,
the solution $d_{r,n}^{\star}$ can be directly given as
\begin{align}
d_{r,n}^{\star}=
\left\{
\begin{aligned}
&d_{r,n,low},\ d_{r,n,sym} < d_{r,n,low}\\
&d_{r,n,up},\ d_{r,n,sym} > d_{r,n,up}\\
&d_{r,n,sym},\ d_{r,n,low} \leq\! d_{r,n,sym}\! \leq d_{r,n,up}
\end{aligned}
\right.
\end{align}

\item[] \underline{CASE-III}:
when
$a_{7,n}\! <\! 0$,
the solution $d_{r,n}^{\star}$ is obtained as
\begin{align}
d_{r,n}^{\star}=
\left\{
\begin{aligned}
&d_{r,n,up},\ d_{r,n,sym} < d_{r,n,low}\\
&d_{r,n,low},\ d_{r,n,sym} > d_{r,n,up}\\
&\textrm{arg} \ \textrm{min}\{ f(d_{r,n,low}),f(d_{r,n,up}) \} 
\end{aligned}
\right.
\end{align}
\end{itemize}
\end{theorem}

Moreover, 
the overall algorithm to solve (P1) is demonstrated in Alg. \ref{alg:Overall_algorithm}.
\begin{algorithm}[t]
\caption{Overall Algorithm to Solve $(\mathrm{P1})$}
\label{alg:Overall_algorithm}
\begin{algorithmic}[1]
\STATE initialize $i=0$;
\STATE randomly generate feasible
$ \mathbf{W}^{0}$,
$\{\mathbf{u}_{k}^{0}\}$,\
$\{q_k^{0}\}$,\
and
$\mathbf{d}_{r}^{0}$;
\REPEAT
\STATE update
$\{\gamma_k\}$,
and
$\{\omega_k\}$
by (\ref{FP_variable_1}) and (\ref{FP_variable_2}), respectively;
\STATE  update $\mathbf{W}^{i+1}$ by Alg. \ref{alg:PDD};
\STATE  update $\{\mathbf{u}_{k}^{i+1}\}$ by equation (\ref{UU_closed_solution});
\STATE  update $\{{q}_{k}^{i+1}\}$ by equation (\ref{p_star});
\STATE  update $\{{d}_{r,n}^{i+1}\}$ by Theorem \ref{Throrem_1};
\STATE $i++$;
\UNTIL{$convergence$}
\end{algorithmic}
\end{algorithm}

\section{Numerical Results}

In this section,
we conduct numerical results to evaluate
the effectiveness of our proposed algorithm.
We assumed that
the BS equipping with
$N_t = N_r = 4$ transmit/receive antennas
simultaneously serves $K=2$ UL users and estimates the DoA of the target.
The BS-target links are
modeled as LoS channels.
Moreover,
we adopt 
a geometric channel model
\cite{ref_channel model_1}
for UL users,
wherein the number of transmit and receive channel paths
is consistently identical, 
i.e.,
$L^{t}_{k} = L^{r}_{k} \triangleq L_1 = 10$.
Thus, 
the path-response matrix for each user is diagonal,
i.e.,
the path-response matrix between the BS and  $k$-th user
is given by
$\mathbf{\Sigma}_{k} = \textrm{diag}\{\sigma^{0}_{k,1}, \dots, \sigma^{0}_{k,L_1}  \}$,
where each
$\mathbf{\sigma}^{0}_{k,l}$ satisfying 
$\sigma^{0}_{k,l} \sim 
\mathcal{CN}(0, {C_0d_{k}^{-\alpha_{loss}}}/{L_1}), 
l= 1, \dots, L_1 $,
where $C_0$  corresponds to the path loss at 
the reference distance of $1$ m,
$d_{k}$ is the propagation distance 
between the BS and the $k$-th user,
$\alpha_{loss} = 2.8$ is the path-loss exponent.
The transmit power of BS is set as $20$dBm. 
The noise power is set as
$\sigma^2 = -80$dBm \cite{ref_noise}.

\begin{figure}[t]
	\centering
	\includegraphics[width=.50\textwidth]{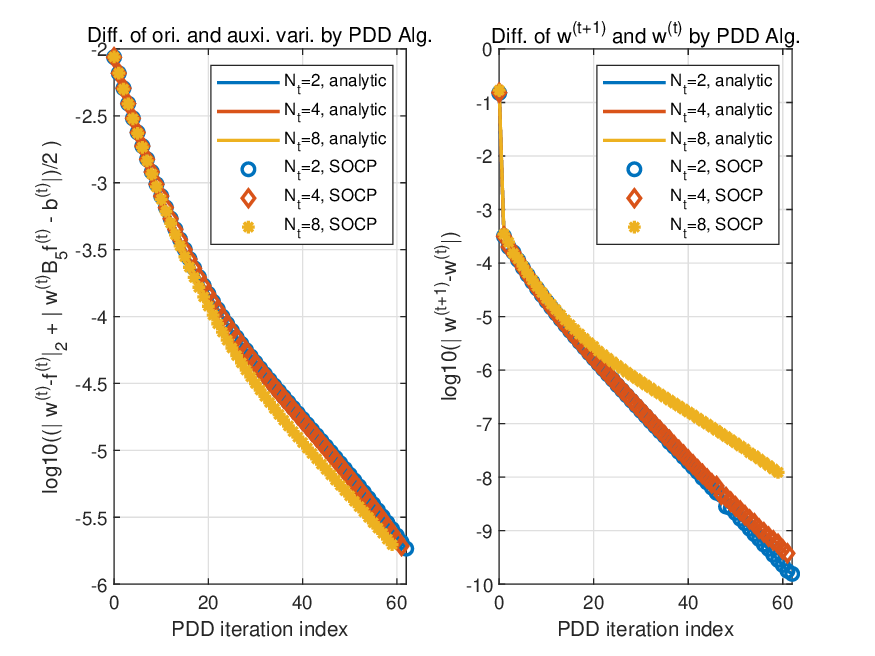}
	\caption{Convergence of PDD method  optimizing $\mathbf{W}$ (Alg. 1).}
	\label{fig.3}
\end{figure}

Fig. \ref{fig.3} 
describes 
the converge behaviors of our
proposed SOCP-based and analytic-based PDD methods
updating the beamformer $\mathbf{W}$.
To ensure a fair comparison, 
both the SOCP and analytic implementations 
start from the same initial point. 
As depicted in Fig. \ref{fig.3},
the left and right subfigures 
respectively display
the differences 
$(\Vert\mathbf{w}^{(t)}-\mathbf{f}^{(t)}\Vert
+ \vert b^{(t)} - (\mathbf{w}^{(t)})^H\mathbf{B}_5 \mathbf{f}^{(t)}\vert)/2 $
and 
$\Vert\mathbf{w}^{(t+1)}-\mathbf{w}^{(t)} \Vert$
in log domain 
 for different settings of the number of BS transmit antenna $N_t$,
 alongside the progression of PDD iterations.
 As indicated by Fig. \ref{fig.3}, 
 both the SOCP and the analytic-based methods achieve nearly identical performance, 
 typically converging well within $60$ iterations.

\begin{figure}[t]
	\centering
	\includegraphics[width=.50\textwidth]{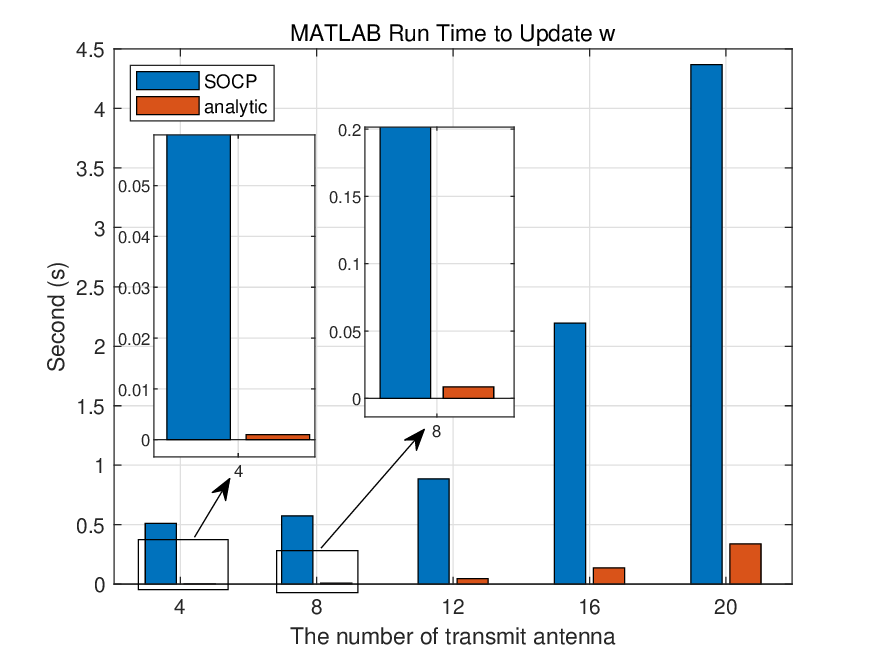}
	\caption{Comparison of time complexity between SOCP and analytical methods for updating $\mathbf{w}$.}
	\label{fig.3_1_1}
\end{figure}

 \begin{figure}[t]
	\centering
	\includegraphics[width=.50\textwidth]{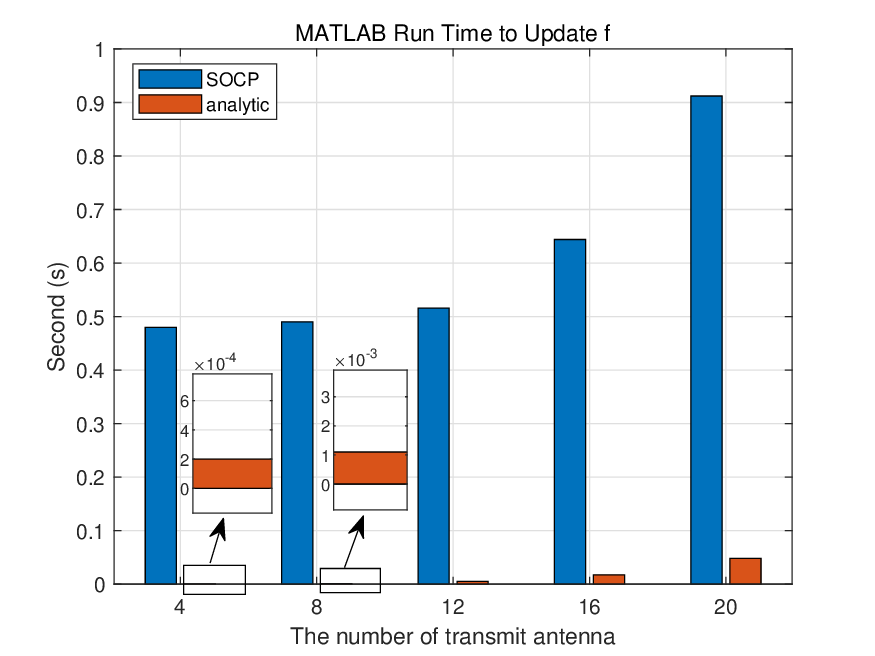}
	\caption{Comparison of time complexity between SOCP and analytical methods for updating $\mathbf{f}$.}
	\label{fig.3_1_2}
\end{figure}
 
  \begin{figure}[t]
	\centering
	\includegraphics[width=.50\textwidth]{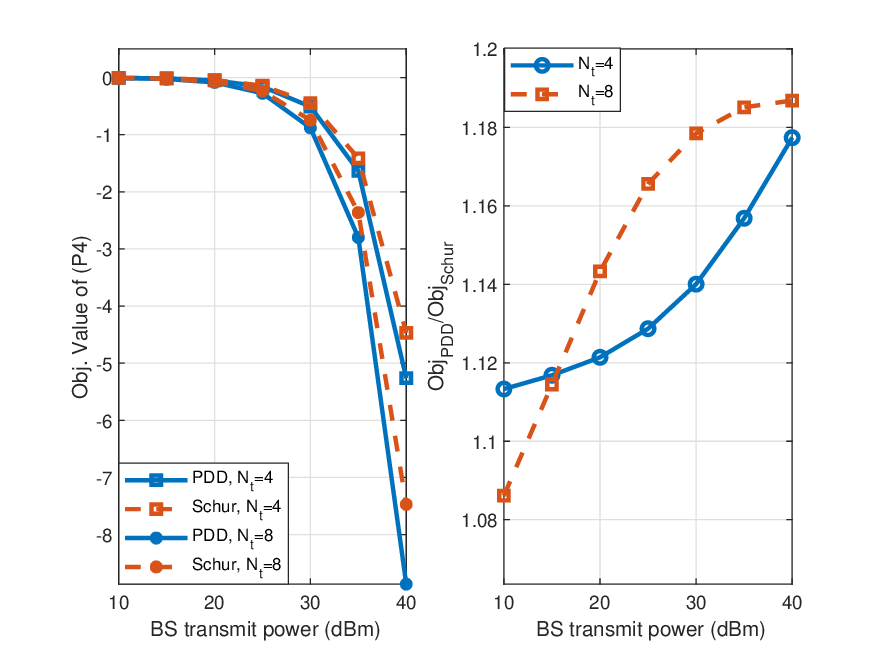}
	\caption{Comparison of performance between PDD-based and Schur-based methods.}
	\label{fig.3_1}
\end{figure}

  \begin{figure}[t]
	\centering
	\includegraphics[width=.50\textwidth]{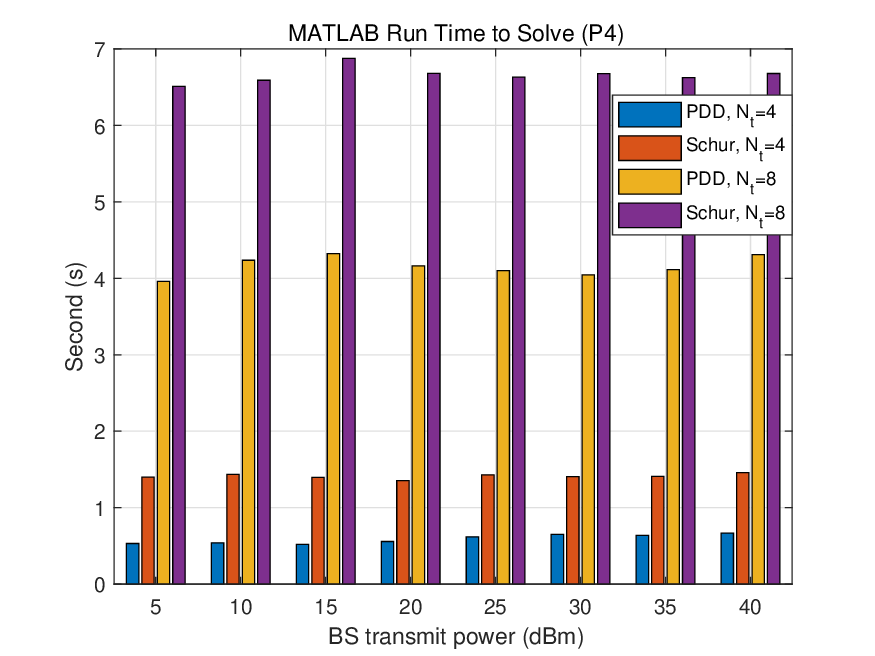}
	\caption{Comparison of time complexity between PDD-based and Schur-based methods.}
	\label{fig.3_2}
\end{figure}


Furthermore, 
we explore the complexity of our proposed analytic-based PDD algorithm 
under various $N_t$ settings. 
The MATLAB runtime comparisons between 
the SOCP-based and analytic-based methods 
are detailed in Fig. \ref{fig.3_1_1} and Fig. \ref{fig.3_1_2}. 
It is important to note that these methods 
achieve identical performance, 
as illustrated in Fig. \ref{fig.3}. 
Moreover,
Fig. \ref{fig.3_1_1} and Fig. \ref{fig.3_1_2}
clearly show that our analytic solutions 
are  highly efficient.
The runtime of the analytic-based method is generally 
two or three orders of magnitude 
less than that of the SOCP-based method.

Firstly, we label the  proposed PDD-based algorithm as ``PDD''.
To verify the performance and time complexity of the proposed PDD-based method for solving (P3),
we introduce the Schur complement proposed in \cite{ref_Convex Optimization} as a benchmark, labeling it as ``Schur''.
In Fig. \ref{fig.3_1}, 
we evaluate the performance of the  PDD-based and Schur-based  algorithms 
across varying numbers of transmit antennas versus BS transmit power. 
The left and right sub-figures correspond respectively to the objective value for solving problem (P3) 
and the ratio of the objective values between the PDD-based and Schur-based algorithms, 
i.e., $\text{Obj}_{\text{PDD}}/\text{Obj}_{\text{Schur}}$. 
The left sub-figure illustrates that 
the curves for all cases decrease as the BS transmit power $P_{BS}$ increases.
Moreover, 
the performance achieved by the PDD-based method closely approximates that of the Schur-based method, 
thereby affirming the efficacy of the PDD-based approach.

Fig. \ref{fig.3_2} illustrates the computation time of the  PDD-based and Schur-based algorithms across varying numbers of transmit antennas.
It is evident that the computation time required by the PDD-based approach is significantly lower than that of the Schur-based approach.
In the case with ``$N_t = 4$'' transmit antennas, 
the computation time of the PDD-based method is on average reduced by 59\% compared to the Schur-based method.
Similarly, 
in the ``$N_t = 6$'' case, 
the computation time of the PDD-based algorithm is reduced by 38\%.
Moreover, combining the data from Fig. \ref{fig.3_1} and Fig. \ref{fig.3_2} allows us to verify the effectiveness and efficiency of the PDD-based method.

  \begin{figure}[t]
	\centering
	\includegraphics[width=.50\textwidth]{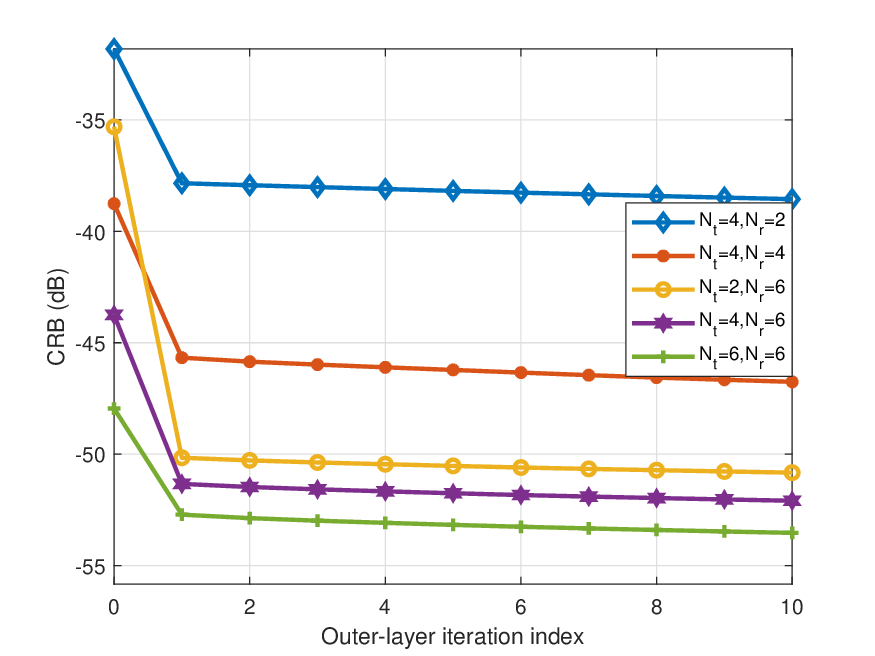}
	\caption{Convergence of Alg. \ref{alg:Overall_algorithm}.}
	\label{fig.4}
\end{figure}

  \begin{figure}[t]
	\centering
	\includegraphics[width=.50\textwidth]{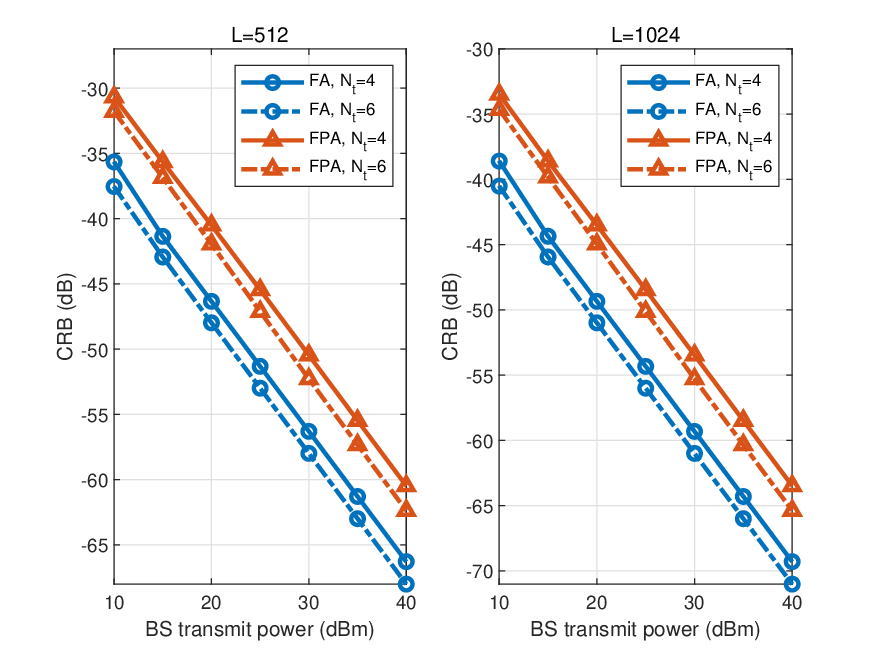}
	\caption{The impact of BS transmit power.}
	\label{fig.5}
\end{figure}

  \begin{figure}[t]
	\centering
	\includegraphics[width=.50\textwidth]{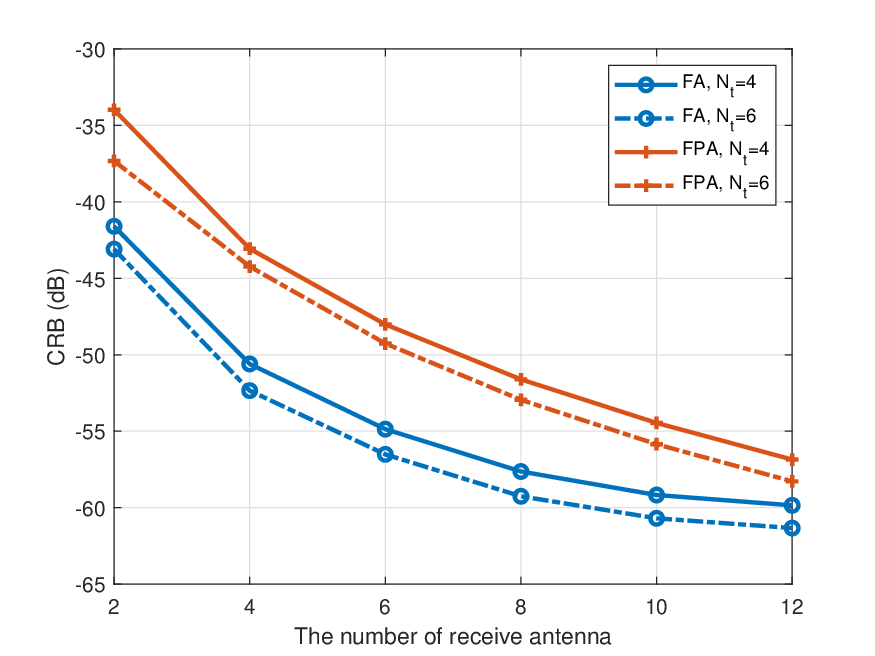}
	\caption{The impact of BS receive antenna number $N_r$.}
	\label{fig.6}
\end{figure}


In Fig. \ref{fig.4},
we present the overall convergence performance of the proposed 
algorithm to tackle the problem (P1) with different BS transmit and/or receive antenna number requirements.
As seen from Fig. \ref{fig.4},
the proposed algorithm can achieve monotonic decrease in CRB
and 
generally achieves significant beamforming gain within around 10 outer iterations.

For comparison,
we consider the following two cases: 
1)``FA'': Our proposed algorithm (i.e., Alg. \ref{alg:Overall_algorithm}) in Sec. III.
2) ``FPA": 
both transmit and receive antennas of BS are equipped with FPA-based array.
Fig. \ref{fig.5} describes the relationship 
between the BS transmit power $P_{BS}$ and CRB.
The left and right subplots correspond to different
numbers of time slot $L$, respectively. 
In this test, 
the transmit power varies from 10dBm to 40dBm.
The CRB of both the ``FA'' and ``FPA'' schemes decreases 
significantly as the transmit power increases, 
and the FA-based scheme improves substantially 
the estimated precision compared to the FPA-based case.
Besides,
compared to the cases with ``L=512'' and ``L=1024'', 
a longer time slot will yield a greater estimated gain.

Fig. \ref{fig.6} depicts
the impact of increasing the number of receive antenna at the BS.
Initially,
it is observed that as the number of receive antenna $N_r$ increases from 2 to 12,
there is a consistent decrease in the CRB across all schemes.
This improvement can be attributed to the additional DoFs provided 
by the increased number of receive antenna.
Then,
the scheme employing FA significantly outperforms the FPA one.
 This superior performance is expected as FA are strategically positioned 
 to modify the channel conditions between the BS and the target, 
 thus enhancing estimation accuracy.
Additionally, 
the CRB for the ``FA'' scenario reaches saturation more quickly than 
that for the ``FPA'' scheme, 
indicating a quicker limit to performance gains with increased receive antenna numbers.

\section{Conclusions}
This paper explores the joint design of  active beamforming and position coefficients in an FA-assisted uplink ISAC system,
which simultaneously performs target angle estimation and uplink communication. 
We propose a toolbox-free and low-complexity solution to jointly design BS probing beamforming, users' power allocation, receiving processors and FA position coefficients.
This approach aims to enhance both target estimation and communication functionalities.
Numerical results present the efficiency and effectiveness of our proposed algorithm and 
highlight the advantages of deploying FA in the uplink ISAC system.

\appendix
\subsection{The Derivation of The FIM in (\ref{F_O_expression})}
\normalem
Firstly,
by vectorizing (\ref{YY_r}), 
we will have
 \begin{align}
 \mathbf{a}_1 = \alpha \mathrm{vec}(\mathbf{A}(\theta_{0})\mathbf{W}\mathbf{S}_r)
 +
 \mathbf{n}_1,
 \end{align}
where 
$\mathbf{n}_1 = \mathrm{vec}(\mathbf{N})
 \sim \mathcal{CN}(0, \sigma^2\mathbf{I}_{N_tL}) $.
 Furthermore,
 the derivatives of 
 $\mathbf{a}_1$ w.r.t.
 $\theta_0$
 and
 $\alpha$
 can be formulated as
  \begin{align}
  &\frac{\partial \mathbf{a}_1}{\partial \theta_0 }
  =
  \alpha \mathrm{vec}(\dot{\mathbf{A}}(\theta_{0})\mathbf{W}\mathbf{S}_r),\\
  & \frac{\partial \mathbf{a}_1}{\partial \bm{\tilde{\alpha}} }
  =
  [1,j]
  \otimes \mathrm{vec}({\mathbf{A}}(\theta_{0})\mathbf{W}\mathbf{S}_r),
  \end{align}
  respectively,
where
 $\dot{\mathbf{A}}(\theta_0)$
 denotes 
 the derivative of 
the cascade channel 
$\mathbf{A}(\theta_0)$ w.r.t.
$\theta_0$,
which is given as
 \begin{align}
 \dot{\mathbf{A}}(\theta_0)
 \triangleq
 \frac{\partial\mathbf{A}(\theta_0) }{\partial\theta_0}
 =
 \dot{\mathbf{a}}_r(\theta_{0})
 \mathbf{a}_t^H(\theta_{0})
 +
 \mathbf{a}_r(\theta_{0})
 \dot{\mathbf{a}}_t^H(\theta_{0}),
 \end{align}
 with
 $\dot{\mathbf{a}}_t(\theta_{0})$
 and
 $\dot{\mathbf{a}}_r(\theta_{0})$
 being defined as
 the
derivatives of 
$ \mathbf{a}_t(\theta_{0})$
 and  
$ \mathbf{a}_r(\theta_{0})$
 with  
 $\theta_{0}$ as follows in (\ref{Appendix_a_1})$-$(\ref{Appendix_a_2}),
 \begin{figure*}
\begin{align}
& \dot{\mathbf{a}}_t(\theta_{0})
 \triangleq
 \frac
 {\partial \mathbf{a}_t}{\partial\theta_{0}}\label{Appendix_a_1}\\
 &=
 \beta_t[j\frac{2\pi}{\lambda}d_{t,1}\cos(\theta_{0})
 e^{j\frac{2\pi}{\lambda}d_{t,1}\sin(\theta_{0})},
 j\frac{2\pi}{\lambda}d_{t,2}\cos(\theta_{0})
 e^{j\frac{2\pi}{\lambda}d_{t,2}\sin(\theta_{0})},
 \cdots,
  j\frac{2\pi}{\lambda}d_{t,N_t}\cos(\theta_{0})
 e^{j\frac{2\pi}{\lambda}d_{t,N_t}\sin(\theta_{0})}
 ]^T,\nonumber\\
 & \dot{\mathbf{a}}_r(\theta_{0})
 \triangleq
 \frac
 {\partial \mathbf{a}_r}{\partial\theta_{0}}\label{Appendix_a_2}\\
 &=
 \beta_r[j\frac{2\pi}{\lambda}d_{r,1}\cos(\theta_{0})
 e^{j\frac{2\pi}{\lambda}d_{r,1}\sin(\theta_{0})},
 j\frac{2\pi}{\lambda}d_{r,2}\cos(\theta_{0})
 e^{j\frac{2\pi}{\lambda}d_{r,2}\sin(\theta_{0})},
 \cdots,
  j\frac{2\pi}{\lambda}d_{r,N_r}\cos(\theta_{0})
 e^{j\frac{2\pi}{\lambda}d_{r,N_r}\sin(\theta_{0})}
 ]^T.\nonumber
\end{align}
\bm{\hrule}
\end{figure*}
 respectively.

The FIM extracted from the observed data (\ref{YY_r})
is given by \cite{ref_FIM_1}
\begin{align}
[\mathbf{F}]_{i,j} 
= 
-\mathbb{E}\bigg[ \frac{\partial^2 \mathrm{In}(f(\mathbf{Y}_r\vert\bm{\zeta} )) }
{\partial {\zeta}_i {\zeta}_j }
  \bigg],
  i, j \in \{1,2,3\}.
\end{align}
Then, the elements
of the FIM 
$\mathbf{F}$
can be expressed in (\ref{Appendix_FIM_1})-(\ref{Appendix_FIM_3}).
\begin{figure*}
\begin{align}
&\mathrm{F}_{\theta_{0}\theta_{0}}\label{Appendix_FIM_1}\\
&
=
\frac{2}{\sigma^2}
\Re\{
\big(\alpha \mathrm{vec}(\dot{\mathbf{A}}(\theta_{0})\mathbf{W}\mathbf{S}_r)
\big)^H
\alpha \mathrm{vec}(\dot{\mathbf{A}}(\theta_{0})\mathbf{W}\mathbf{S}_r)
\}
=
\frac{2\vert\alpha \vert}{\sigma^2}
\Re\{
\mathrm{tr}
(
\dot{\mathbf{A}}(\theta_{0})\mathbf{W}\mathbf{S}_r
\mathbf{S}_r^H\mathbf{W}^H\dot{\mathbf{A}}^H(\theta_{0})
)
\}\nonumber\\
&=
\frac{2L\vert\alpha \vert}{\sigma^2}
\mathrm{tr}
(
\dot{\mathbf{A}}(\theta_{0})\mathbf{W}\mathbf{W}^H\dot{\mathbf{A}}^H(\theta_{0})
)
=
\mathrm{tr}(
\dot{\mathbf{A}}^H(\theta_{0})
\dot{\mathbf{A}}(\theta_{0})
\mathbf{W}\mathbf{W}^H
),\nonumber\\
&\mathbf{F}_{\theta_{0}\alpha}\label{Appendix_FIM_2}\\
&=
\frac{2}{\sigma^2}
\Re\{
\big(\alpha \mathrm{vec}(\dot{\mathbf{A}}(\theta_{0})\mathbf{W}\mathbf{S}_r)
\big)^H
[1,j]
\otimes \mathrm{vec}({\mathbf{A}}(\theta_{0})\mathbf{W}\mathbf{S}_r)
\}
=
\frac{2}{\sigma^2}
\Re\{
\big(\alpha^{\ast}
[1,j]
\mathrm{tr}
(
\mathbf{S}_r^H\mathbf{W}^H\dot{\mathbf{A}}^H(\theta_{0})
{\mathbf{A}}(\theta_{0})\mathbf{W}\mathbf{S}_r
)
\}\nonumber\\
&=
\frac{2L}{\sigma^2}
\Re\{
\big(\alpha^{\ast}
[1,j]
\mathrm{tr}
(\mathbf{W}^H\dot{\mathbf{A}}^H(\theta_{0})
{\mathbf{A}}(\theta_{0})\mathbf{W}
)
\}
=
\frac{2L}{\sigma^2}
\Re\{
\big(\alpha^{\ast}
[1,j]
\mathrm{tr}
(
\dot{\mathbf{A}}^H(\theta_{0})
{\mathbf{A}}(\theta_{0})\mathbf{W}\mathbf{W}^H
)
\},\nonumber\\
&\mathbf{F}_{\alpha\alpha}\label{Appendix_FIM_3}\\
&=
\frac{2}{\sigma^2}
\Re\{
\big([1,j]
\otimes
\mathrm{vec}({\mathbf{A}}(\theta_{0})\mathbf{W}\mathbf{S}_r)
\big)^H
[1,j]
\otimes \mathrm{vec}({\mathbf{A}}(\theta_{0})\mathbf{W}\mathbf{S}_r)
\}
=
\frac{2}{\sigma^2}
\Re\{
[1,j]^H[1,j]
\mathrm{tr}
(
{\mathbf{A}}(\theta_{0})\mathbf{W}\mathbf{S}_r
\mathbf{S}_r^H\mathbf{W}^H{\mathbf{A}}^H(\theta_{0})
)
\}\nonumber\\
&=
\frac{2L}{\sigma^2}
\mathrm{tr}
(
{\mathbf{A}}(\theta_{0})\mathbf{W}
\mathbf{W}^H{\mathbf{A}}^H(\theta_{0})
)
=
\frac{2L}{\sigma^2}
\mathrm{tr}
(
{\mathbf{A}}^H(\theta_{0}){\mathbf{A}}(\theta_{0})
\mathbf{W}
\mathbf{W}^H)\mathbf{I}_2.\nonumber
\end{align}
\bm{\hrule}
\end{figure*}

\subsection{Proof of (\ref{MA_BS_dd})}
\normalem
Proof:
Since the feasible set of the problem (P27) is existing,
the equation 
$\tau_3d_{r,n}^2
+b_{14,n}d_{r,n} + c_{26,n}
= 0$ 
possesses at least one solution.
Firstly,
if the above equation only has one solution, i.e., $d_{r,n,r1}=d_{r,n,r2} \in [d_{r,n-1} +d_{min}, d_{r,n+1}-d_{min}] $,
we can directly set $d_{r,n}^{\star} =d_{r,n,r1} $.

Furthermore,
if the equation 
$\tau_3d_{r,n}^2
+b_{14,n}d_{r,n} + c_{26,n}
= 0$ has two roots, i.e., $d_{r,n,r1} < d_{r,n,r2}$,
the feasible area of (P27)
can be determined by the following two cases:

\underline{CASE-1}:
When $ d_{r,n-1} +d_{min} < d_{r,n,r1}$,
we have two sub-cases:

{case-\Circled{1}}:
If $d_{r,n,r1} \leq d_{r,n+1}-d_{min} < d_{r,n,r2}  $,
the feasible set is $[d_{r,n,r1},   d_{r,n+1}-d_{min}]$.

{case-\Circled{2}}:
If $d_{r,n,r2} \leq d_{r,n+1}-d_{min}  $,
the feasible set is expressed as
 $[d_{r,n,r1},   d_{r,n,r2}]$.

\underline{CASE-2}:
When 
$ d_{r,n,r1} \leq d_{r,n-1} +d_{min}  $,
we have the following two sub-cases:

{case-\fbox{1}}:
If $  d_{r,n+1}-d_{min}  \leq d_{r,n,r2} $,
the set $[d_{r,n-1} +d_{min} ,d_{r,n+1}-d_{min} ]$ is feasible set.

{case-\fbox{2}}:
If $     d_{r,n,r2} \leq d_{r,n+1}-d_{min} $,
 $[d_{r,n-1} +d_{min}, d_{r,n,r2} ]$
 is a feasible area.

Based on \underline{CASE-1} and \underline{CASE-2},
we can obtain the feasible area of (P27) as follows:
\begin{align}
d_{r,n}\in [ &\textrm{max}\{ d_{r,n,r1}, d_{r,n-1} +d_{min} \},\\
&\textrm{min}\{d_{r,n,r2}, d_{r,n+1}-d_{min}
\}
 ].\nonumber
\end{align}


\end{document}